\documentclass[preprint2]{aastex701}
\usepackage{amsmath}
\usepackage{xcolor}

\newcommand{\resubi}[1]{#1}
\newcommand{\resubii}[1]{#1}

\begin{document}

\title{Understanding how Coronal Heating and Magnetic Field Scaling Affect Global Properties of the Low and Middle Corona}

\author[orcid=0000-0002-6478-3281]{Caroline L. Evans}
\affiliation{Department of Astrophysical and Planetary Sciences, University of Colorado Boulder, 2000 Colorado Avenue, Boulder, CO 80305, USA}
\affiliation{Cooperative Institute for Research in Environmental Sciences, 216 UCB Boulder, CO, 80309, USA}
\affiliation{National Solar Observatory, 3665 Discovery Dr, Boulder, CO 80303, USA}
\email[show]{caroline.evans-1@colorado.edu}

\author[orcid=0000-0003-1759-4354]{Cooper Downs}
\affiliation{Predictive Science Inc., 9990 Mesa Rim Road, Suite 170, San Diego, CA 92121, USA} 
\email{cdowns@predsci.com}  

\author[orcid=0000-0002-9654-0815]{Donald Schmit}
\affiliation{Cooperative Institute for Research in Environmental Sciences, 216 UCB Boulder, CO, 80309, USA}
\email{donald.schmit@colorado.edu}  
 
\begin{abstract}

The magnetic field of the Sun is a key driver of coronal heating and solar wind acceleration. In tandem, these influence the observable morphology of the corona, along with defining the magnetic connectivity and magnitude of open magnetic flux. Here, we investigate how coronal heating affects the magnetothermal state of the 3D global corona by analyzing a series of data-constrained, thermodynamic magnetohydrodynamic (MHD) simulations for solar minimum conditions. We vary the overall heat deposited in the corona via modifying the input Poynting flux of a wave-turbulence-driven (WTD) model for coronal heating and by scaling the inner boundary magnetic field. We quantify the magnetic, thermodynamic, and combined magnetothermal effects on our simulations. We find that increasing the overall amount of heat leads to a higher magnitude of open flux and a more visibly open white-light morphology in the K-Corona between 1.5 and 6 \resubi{R$_\odot$}. We quantify how plasma $\beta$ evolves and drives morphological differences between simulations. We also compare these runs to an extremely high-resolution reference solution that includes small-scale magnetic information in the surface boundary condition. This latter comparison provides insight into the further development of ‘subgrid’ corrections for low(er) resolution models to best represent the structure and observables of the corona and inner heliosphere.

\end{abstract}


\section{Introduction} \label{sec:intro}

The solar magnetic field influences the morphology of the corona and vice versa. The photosphere is the primary source of the magnetic field that extends into the corona. As a highly rarefied plasma, the coronal magnetic field influences the structure of the corona, across spatial scales. Analyzing the magnetic field -- rooted in the photosphere and extending into the corona -- is critical to understand the formation and evolution of the corona, as well as solar wind acceleration. 

The Sun's magnetic field is central to two unresolved problems relevant to the solar wind: coronal heating and open flux. Theories explaining coronal heating typically fall into two classes of proposed mechanisms. Both ``AC'' or ``DC'' mechanisms \citep[see][and references therein]{Klimchuk2006} depend on the constant, roiling motion of the base (or ``footpoints'') of coronal loops to create the Poynting flux that provides the energy source for coronal heating. The resulting heat sets the densities and temperatures of the corona, critical to setting solar wind parameters \citep{Parker1958}. 

The open flux problem is the mismatch between measurements of the photospheric magnetic field from magnetically sensitive spectral lines with in-situ measurements of the interplanetary magnetic field (IMF) \citep{Lowder2017, Linker2017, Wang2022}. Our observational definition of open flux indicates that less dense, cooler coronal holes are the source of open flux \citep{Cranmer2009}. The emission (or lack thereof) of the extreme-ultraviolet (EUV) corona reveals the morphology of the open field. However, \citet{Linker2017} show that coronal hole detection on synthetic observables does not perfectly match the ground truth open flux for a 3D magnetohydrodynamic (MHD) simulation. A key question therefore arises: does heating influence the amount of open flux in the corona?

The photospheric magnetic field is key to the open flux problem, but remote sensing of the photospheric magnetic field makes its precise measurement difficult. Magnetographs often underestimate the magnitude of magnetic flux \citep{Centeno2023, Milic2024}; proposed solutions include scaling the measurements \citep{Petrie2025} or increasing energization of the corona\resubii{. This is seen in the Aly-Sturrock conjecture \citep{Aly1991, Sturrock1991}, indicating that for a corona-like force-free field, the most energetic state is not a highly twisted closed field but rather a fully open field, extending from the surface to infinity -- or as in this case, the inner heliosphere}. The uncertainties of inferring the magnetic field of the photosphere (see \cite{delToroIniesta2016} and references therein for a comprehensive overview of magnetic field inversions)\resubi{, lack of routine measurements of the polar fields \citep{Riley2019},} and in measuring photospheric electric fields and currents \citep{Schuck2022, Titov2025} make definitive conclusions elusive. 

While seemingly unconnected, the solar wind connects coronal heating and open flux as the former sets the plasma pressures and velocities that travel  along the IMF that the latter sets. Further, the morphology and strength of the photospheric magnetic field determines the open flux and the heating of the corona. In the middle corona \citep[as defined in][ ranging from $\sim$ 1.5 - 6 R$_\odot$]{West2023}, this interplay of the magnetic field and coronal heating is apparent. The middle corona marks the transition from predominantly closed to open structures \citep{Seaton2021}, variable density structures \citep{DeForest2018}, and hosts the Alfv\'en surface \citep{Cranmer2023}. Understanding this regime \resubi{would provide} valuable insight into the connection between the low corona and solar wind \citep{ViallBorovsky2020}.

On a more practical level, constructing a data-constrained thermodynamic MHD model of the global solar corona (i.e. one where the thermodynamic transport processes and magnetic field dynamics are simultaneously captured) generally requires two key decisions to be made: which coronal heating prescription to use and what magnetic boundary conditions to use at the base. Different heating models change the morphology of MHD solutions \citep{Lionello2009, Downs2010}. The large-scale coronal solution is also strongly dependent on the magnetic data sources and map construction for the inner boundary condition \citep{Liu2026, Baratashvili2026, Sachdeva2026}. In this study, we systematically investigate both how changing the heating and scaling of the magnetic boundary condition influence various observables and MHD state variables.

Previous work posits that the presence of small-scale structures in the photospheric magnetic field influences the net coronal heating rate \citep{Evans2026}. In a data-driven, 3D thermodynamic global MHD model, they demonstrate that an approximately four times increase in \resubi{angular} resolution leads to 40\% more heating on the smallest and global scales. We use these results to motivate changing the heating rate by (reciprocal) increments of 40\% and scaling of the magnetic field to investigate (i) how and why the resulting simulations change and (ii) if it is possible to replicate the results of the highest resolution model at a lower resolution.

This paper proceeds as follows: in $\S$ \ref{sec:methods} we outline the model central to this work, relevant numerical analysis techniques, and our numerical experiment. In $\S$ \ref{sec:results}, we present \resubi{morphological,} magnetic, thermodynamic, and combined magnetothermal differences between our simulations. We offer conclusions about the role of heating \resubi{and magnetic field strength} in shaping coronal morphology in $\S$ \ref{sec:conclusions}.

\section{Numerical Methods} \label{sec:methods}

\subsection{Magnetohydrodynamic Algorithm outside a Sphere} \label{sec:MAS}

Coronal morphology is indelibly related to its thermodynamics. Therefore, we use the data-constrained, wave-turbulence-driven (WTD) thermodynamic 3D MHD model, Magnetohydrodynamic Algorithm outside a Sphere (MAS). Solutions from MAS model a wide variety of coronal phenomena \citep{Mikic1999, Lionello2009, Riley2011, Lionello2013CME, Torok2018}. In this paper we leverage a series of previous simulations that were performed for modeling the corona during the 2019 total solar eclipse. All simulations presented here are \resubi{nearly identical} to the MAS simulation presented in \resubi{\citet{Boe2021, Boe2022, Boe2023}}, which compared favorably to broad-band white-light observations (K-Corona) and several narrow-band coronal emission lines observed in visible wavelengths.

MAS incorporates a Wentzel-\resubi{Kramers}-Brillouin (WKB) approach to model the wave-pressures responsible for solar wind acceleration and a WTD model for coronal heating \citep[see e.g.][]{Jacques1977}. This combination encompasses linear effects of coronal heating (Alfv\'en wave transport) as well as nonlinear effects (Alfv\'en wave reflection and dissipation). This is the same heating implementation as \cite{Downs2016} (1D) and \cite{Mikic2018} (3D). The Els\"asser variables, where $\mathbf{z}_\pm = \delta \mathbf{v} \mp \delta \mathbf{B}/ \sqrt{4\pi\rho}$, govern the WTD heating implementation. The Els\"asser variables are a flow velocity field ($\mathbf{v}$) perturbation via the Alfv\'en velocity, which depends on the magnetic field ($\mathbf{B}$) and density ($\rho$). The following partial differential equation solves for the Els\"asser variables:

\begin{gather}\label{eqn:heat}
    \frac{\partial z_\pm}{\partial t} + (\mathbf{v} \pm \mathbf{v_A}) \mathbf{\cdot \nabla} z_\pm = R_1 z_\pm + R_2 z_\mp - \frac{|z_\mp| z_\pm}{2\lambda_\perp} \\
    R_1 = \frac{1}{4} (\mathbf{v} \mp \mathbf{v_A}) \mathbf{\cdot \nabla} (\mathrm{log}\rho)\\
    R_2 = \frac{1}{2} (\mathbf{v} \mp \mathbf{v_A}) \mathbf{\cdot \nabla} (\mathrm{log}|\mathbf{v_A}|)
\end{gather}

The advective derivative and $R_1$ term represent linear wave propagation from WKB. The remaining terms represent the non-linear, WTD approach; $R_2$ represents self-reflection and $\lambda_\perp$ is the dissipation term. The \resubi{net} WTD heating term is defined as $H = \frac{\rho}{4 \lambda_\perp} (|z_-|z_+^2 + |z_+|z_-^2)$. $\lambda_\perp$ represents the transverse correlation scale of fluctuations and is defined as $\lambda_0 \sqrt{B_0/B}$, where the constants $\lambda_0$ and reference magnetic field $B_0$ define empirically chosen free parameters of 13.92 Mm and 8.537 G, respectively.

The most important free parameter for WTD heating is the Els\"asser amplitude of the waves propagating outward into the corona from the inner boundary, z$_0$. It specifies the value of $z_+$ when the radial component of the field at the inner boundary is positive, and $z_-$ when it is negative. The combination of constant electron density ($4\! \times \! 10^{12}$ cm$^{-3}$) and $z_0$ at the inner boundary determines the wave energy at the boundary, driving the heating rate for our MHD simulations. 
The resulting Poynting flux, as defined in \citet{Downs2016}, is
\begin{equation} \label{eq:downs_pflux}
    P_\pm = \pm \frac{B \rho^{\frac{1}{2}}z_\pm^2}{4\sqrt{4\pi}} .
\end{equation}

Traditionally, the Poynting vector is the amount of energy per area per time, representing the flow of electromagnetic energy and is most often defined as the cross product between the elementary electric field and the magnetizing field:

\begin{equation}
    S = \frac{c}{4\pi}(\mathbf{E} \times \mathbf{H}) .
\end{equation}

In cgs units, this has units of $\mathrm{erg/cm^2/s}$. This is the same units as Equation \ref{eq:downs_pflux}. 

Whereas simulations or observations targeting spatial scales on the order of kilometers relate the Poynting flux to magnetic energization or helicity injections, a global MHD \resubi{solution simulates larger spatial scales.} Consequently, our Poynting flux is the flux resulting from inherently unresolved ``subgrid'' fluctuations. Here, ``subgrid'' indicates anything occurring on spatial scales smaller than the resolution of the simulation. These subgrid fluctuations still contribute to the MHD equations through contributing energy to the heating rate. \resubi{As seen in Equation \ref{eq:downs_pflux}, the outward and inward Poynting flux depend on the Els\"asser variables z$_\pm$. Further, the net WTD heating rate depends on the solution of the Els\"asser variables.} Thus, the net Poynting flux is the net WTD heating rate\resubi{, incorporating subgrid information.}

Our goal is to vary the input Poynting flux to examine the influence of heating on coronal morphology and the large-scale solar wind morphology (e.g. amount of open flux).
In contrast to \cite{Evans2026}, which explores the effects of resolution on coronal morphology, all new simulations in this work (\S\ref{sec:experiment_design}) have the same resolution. Each of the 288 radial slices have a resolution of 148 \resubii{$\times$ } 315 ($\theta$ \resubii{$\times$ } $\phi$ or Co-Latitude x Longitude). The mesh is uniform in Longitude and within $\pm 65^{\mathrm{o}}$ in Co-Latitude; it is otherwise nonuniform. Unlike \cite{Mason2023, Lionello2023, Downs2025}, these simulations are \resubii{quasi-}steady-state solutions, \resubii{where the simulation is relaxed for a long period of time with a static boundary condition for the radial component of the magnetic field}. \resubii{We relaxed the new simulations in this paper for 80 physical hours, sufficient time for streamers to form and the solar wind to reach our outer radial boundary (30 R$_\odot$).} We do present comparisons to one run from \citet{Evans2026}; the super resolution run. This simulation has a mesh of 445 × 569 × 1257 ($r$ x $\theta$ x $\phi$). \resubi{The super angular resolution is approximately a factor of four better and there are over 25 times more points in it than in the new simulations in this work. We use a lower resolution because that work’s super resolution simulation is cost-prohibitive to routinely run.} This \resubi{set} of simulations allows us to (i) isolate the effects on morphology, magnetic properties, and thermodynamic effects with intra-resolution comparison and (ii) test the best way to emulate a simulation that incorporates high\resubi{-}resolution surface magnetic fields with a more modest simulation setup and grid that fundamentally cannot.

\subsection{Relevant Numerical Analysis Techniques} \label{sec:a_techniques}

\subsubsection{Forward Modeling} \label{sec:forwardmodel}

Synthetic observables allow for morphology comparisons that are otherwise challenging for a 3D simulation. To present accurate synthetic images of the corona, we perform ``forward modeling.'' We use the state variables from our MHD solutions to estimate the emissivities of various observables and produce integrated, line-of-sight white-light or EUV images. 

We synthesize observables most similar to the Solar Dynamics Observatory/Atmospheric Imaging Assembly \citep[SDO/AIA][]{Lemen2012} given its widespread use. For emission lines in the EUV, we use a technique similar \resubi{to} \citet{Lionello2009} while also accounting for relevant opacities in the same manner as \citet{Mok2016}. For each line-of-sight (LOS), $s$, emission can be calculated according to:

\begin{equation}
    I = \int_{LOS} n_e^2 \ R(n_e, \ T_e) \ e^{-\tau_i(s)} \ ds,
    \label{eq:emission}
\end{equation}

where this is a LOS integral (ds) over the electron density ($n_e$) and a given response function ($R$), which depends on $n_e$ and electron temperature ($T_e$). The exponential term describes extinction through opacity in such a system, which is determined by the integrated column density of absorbers. For the case of continuum absorption by cold material \citep[e.g. for species H I, He I, and He II \resubi{given in}][]{AnzerHeinzel2005}, it depends on the optical depth $\tau$, which is defined as
 \begin{equation}
     \tau_i(s) = \sigma_i \int_{-\infty} ^{s} n_i \ ds'.
 \end{equation}

Here, $\sigma$ represents the absorption cross section for a species of interest and $n_i$ is the number density of that species. This allows for the correct extinction of emission for LOSs that intersect the model chromosphere just off the limb.
For some stronger coronal emission lines, we include a heuristic treatment for resonant scattering at the base of the corona and transition region \citep{Mok2016}, which is described in $\S$ \ref{sec:appendix}. 

For our white-light images, we compute the K-Corona polarization brightness (pB) as determined by the \resubi{Thomson} scattering of broadband photospheric emission by coronal electrons. Many ground- and space-based coronagraphs routinely capture pB as it encapsulates the overall morphology of the global corona. Furthermore, the overall brightness of pB is directly proportional to the coronal electron density ($n_e$) making it a useful absolute density diagnostic. We follow the procedure of \citet{Downs2025}, which follows the method provided in \citet{HowardTappin2009}, to generate pB.

\subsubsection{Structural Similarity Index (SSIM)} \label{sec:ssim}

We use the structural similarity index \citep[SSIM,][]{Wang2004, Wang2009}, an image processing technique, to test the similarity between simulations with different resolutions.  Inherently, if two simulations have different resolutions, the one with higher resolution will be able to resolve smaller-scale structures, particularly if higher resolution information is included in the boundary condition (as is the case here). These structures would heavily influence traditional image comparison techniques such as mean square error. In contrast, SSIM, as implemented in scikit-image\footnote{\url{https://scikit-image.org/docs/stable/auto\_examples/transform/plot\_ssim.html}}, is designed to replicate human ratings of image similarity \resubi{(holistic comparison)} and is less sensitive to differences in resolution. SSIM has been used extensively in solar physics image comparison, including for machine learning \citep[][]{Jarolim2024, Grajeda2025, Jarolim2025, Schirninger2025}. A value of 1 indicates identical images. This metric represents an opportunity to test for similarity between images of different resolutions, with the aim to determine how well a lower-resolution run replicates the results of a higher-resolution run. We apply it to compare our synthetic white-light and EUV images of our new simulations to those of our higher-resolution reference simulation.

\resubi{As SSIM compares images, it is important to scale scientific data products similarly (or in this case, identically) to compare them on the same basis. For each type of image comparison (EUV or white-light), we use the same number of pixels in x and y. The same data minimum, maximum, color table, and normalization scheme (log-space) is also used to rasterize each image. The EUV images have a field of view extending to $\pm$1.4 R$_\odot$ and the white light images extend to $\pm$4 R$_\odot$. The white light images are radially filtered using the same fixed radial function, whereas the EUV images are not. Also, for EUV images, we do not mask any off-limb or on-disk regions because we are interested in seeing morphological differences, both on- and off-disk. The higher-resolution reference simulation is rendered the same way as the other simulations to provide the data range to SSIM to help standardize our results. }

\subsubsection{Spherical Harmonic Decomposition (SHD)} \label{sec:shd}

Given the cross-resolution analysis inherent to this work, we use spherical harmonic decomposition (SHD) to target specific spatial scales. As in \citet{Evans2026}, we quantify the spatial scales of the coronal heating rate in these MHD solutions \citep{Wieczorek2018}. We also utilize total power spectra (TPS) as our evaluative metric. TPS depend on the degree $\ell$ (and not the order $m$) of the spherical harmonic, indicating the quantity of structures at a given spatial scale. The physical length scale is inversely proportional to the angular distance $\ell$ and consequently is also a function of the radial distance from the photosphere. Therefore, the largest spatial scales correspond to the smallest $\ell$-values (and vice versa).  We also use the closure offered by Parseval's theorem, which states that the sum of the total power spectrum is the integral of the square of a quantity of interest. In practice, we take the square root of the absolute value of the heating rate to produce power spectra with units of erg/(cm$^3$ s).

\subsubsection{Plasma $\beta$}

The dimensionless plasma $\beta$ represents the ratio between the gas pressure to magnetic pressure: \resubi{ $\frac{\ n \ k_B T}{B^2/(8\pi)},$ where $n$ is the total number density of particles. The MHD simulation setup used here assumes that the plasma consists only of protons and electrons, and that the ion and electron temperatures are equal ($T_i = T_e$).}  

\resubi{$\beta$} roughly indicates when the plasma pressure, a hydrodynamic quantity, more heavily influences a regime ($\beta > 1$), versus when the magnetic forces shape a region ($\beta < 1$). The middle corona is a critical region for plasma $\beta$ transitions for the quiet Sun\resubi{; it further} is a relevant parameter for coronal holes and streamer regions alike \citep{West2023}, making its inclusion in this work key. 

\subsection{Simulation Experiment Design}
\label{sec:experiment_design}
 
MHD models have adjustable parameters to tune their output. The results of \citet{Evans2026} found that \resubi{$\sim$40\%} of the net heating in the low corona in an extremely high resolution run was due to small-scale magnetic field elements that were inherently missing in lower resolution equivalents. Using this result as a benchmark, we vary the WTD parameter z$_0$ to change the amount of heating. We set values of z$_0$ \resubi{using} reciprocal spacing: 1/1.8, 1/1.4, 1.4, and 1.8. This results in our simulations of 44\% less (44l), 29\% less (29l), 40\% more (40m), and 80\% more (80m) outward Poynting flux. \resubi{The typical value of z$_0$ is 9.627 km/s, as we use in ``base'' and ``super.'' In Equation \ref{eq:downs_pflux}, the Poynting flux depends on z$_0^2$. As a result, the z$_0$ value required to yield 40\% more heating is the product of 9.627 and $\sqrt{1.4}$. Similarly, 29l results from scaling the base 9.627 km/s by $\sqrt{1/1.4}$.} Reference solutions where we do not change the z$_0$ parameter are referred to as a ``base'' simulation. We lastly include the extremely high resolution run, the ``super'' resolution simulation, described in \citet[][]{Evans2026} as a reference solution. 

\resubii{We note that because the super run represents a significant computational expense, we did not create an unscaled ``super'' resolution equivalent. Along similar lines, we omit the unscaled 29\% less simulation because the trends are clear from the existing simulations.}

We examine how the net Poynting flux influences the morphology of the corona by modifying two different parameters. As in Equation \ref{eq:downs_pflux}, modifying z$_0$ changes the net Poynting flux in these simulations. However, the net Poynting flux also depends on the strength of the magnetic field. In addition to scaling z$_0$, we either scale the magnetic \resubi{field} by a factor of 1.4 or leave this unscaled. As a result, we have \resubi{created} two sets of parallel simulations: Set S (S: for scaled) and Set U (U: for unscaled)\resubi{, summarized below in Table \ref{tab:sum}}.

\begin{deluxetable*}{cccccc}[hbt!]
\label{tab:sum}
\tablecolumns{6}
\tablewidth{0pt}
\tablecaption{Simulations Summary} 
\tablehead{\colhead{Set S, scaled} & 
\colhead{\shortstack{44\% less \\ (S: 44l)}} & 
\colhead{\shortstack{29\% less \\ (S: 29l)}} & 
\colhead{\shortstack{base/super\\ (S: base/S: super)}} & 
\colhead{\shortstack{40\% more\\ (S: 40m)}} & 
\colhead{\shortstack{80\% more \\ (S: 80m)}}}
\startdata
z$_0$ value (km/s) & 7.158 & 8.126 & 9.627/ 9.627  & 11.408 & 12.949 \\
Net Poynting Flux (10$^{28}$ erg/s) & .572 & .736 &  1.03/1.57 & 1.46 & 1.89 \\
\hline 
\hline 
  & 44\% less  &  -  &  base   & 40\% more    &  80\% more   \\
 Set U, unscaled & (U: 44l)  &  -  &  (U: base)  & (U: 40m)   &   (U: 80m)   \\
\hline
z$_0$ value (km/s)  & 7.158  &  -  & 9.627  & 11.408 & 12.949 \\
Net Poynting Flux (10$^{28}$ erg/s)  &  .393  & - & .714   & 1.01  & 1.31  
\enddata
\end{deluxetable*}

We present a quantitative summary of our experiment in Table \ref{tab:sum}. The top half of the table indicates the corresponding values for Set S and the bottom half does the same for Set U. The top row indicates the value of z$_0$, the Els{\"a}sser variable magnitude at the photosphere (1 R$_\odot$) in km/s (as defined in $\S$ \ref{sec:MAS}). \resubi{ We then }show the net Poynting flux through the inner boundary in $10^{28}$ erg/s (Equation \ref{eq:downs_pflux}). This differs from the specified, outward Poynting flux because it accounts for losses of wave energy back through the inner boundary either through self-reflection or propagation to the opposite end of the loop. This net value depends on the 3D plasma and magnetic field properties of the MHD simulation and, through energy conservation, represents the integrated heating of the WTD model in the volume.

\resubi{Our net Poynting flux values are in reasonable agreement with both modeling and observational cases. 
\citet{Huang2023, Huang2024} vary the input Poynting flux to better match key solar wind parameters over the solar cycle. Scaling the values in \citet{Huang2023, Huang2024} over the full-Sun-area, they respectively report $2.86 \times 10^{28}$ erg/s, and $3.16 \times 10^{28}$ erg/s. As  our simulation is at solar minimum, it is reasonable that these values exceed all of ours, including our most heated, scaled case (S: 80m).}

\resubi{Observationally, \citet{Morton2025} report the average Poynting flux in polar coronal holes over the solar cycle. Again scaling for our global domain, this value is $6.02 \times 10^{27}$ erg/s. Only in cases of under-heating (S: 44l and S:29l) and under-scaling (U: 44l and U: base) of our solution are the values similar. It is unsurprising that this value is on the lower end of our solar minimum simulations given that coronal holes produce lower Poynting flux than the global Sun given their lower densities.} 

The net \resubi{Poynting} flux increases, as expected, with z$_0$. The amount of net Poynting flux is an inherently nonlinear quantity given its dependence on z$_{\pm}^2$. The mismatch between simulations that have similar initial Poynting fluxes encapsulates this change: S: base \& U: 40m and U: 80m \& S: 40m. However, it is not just the Poynting flux that varies between these simulations. Modifying the amount of heating via z$_0$ alters the densities and changing the scaling of the magnetic field influences the effective plasma $\beta$ in the middle corona (see $\S$ \ref{sec:beta}).

\section{Results} \label{sec:results}

In this section, we discuss the various ways in which the magnetic field and hydrodynamic properties change in our experiment. We quantify how the models change under our net Poynting flux modifications and evaluate if any such parameter space replicates the results of our higher resolution reference solution.

\subsection{Morphological Effects} \label{sec:morph}
\begin{figure*}
    \centering
    \includegraphics[scale=0.88]{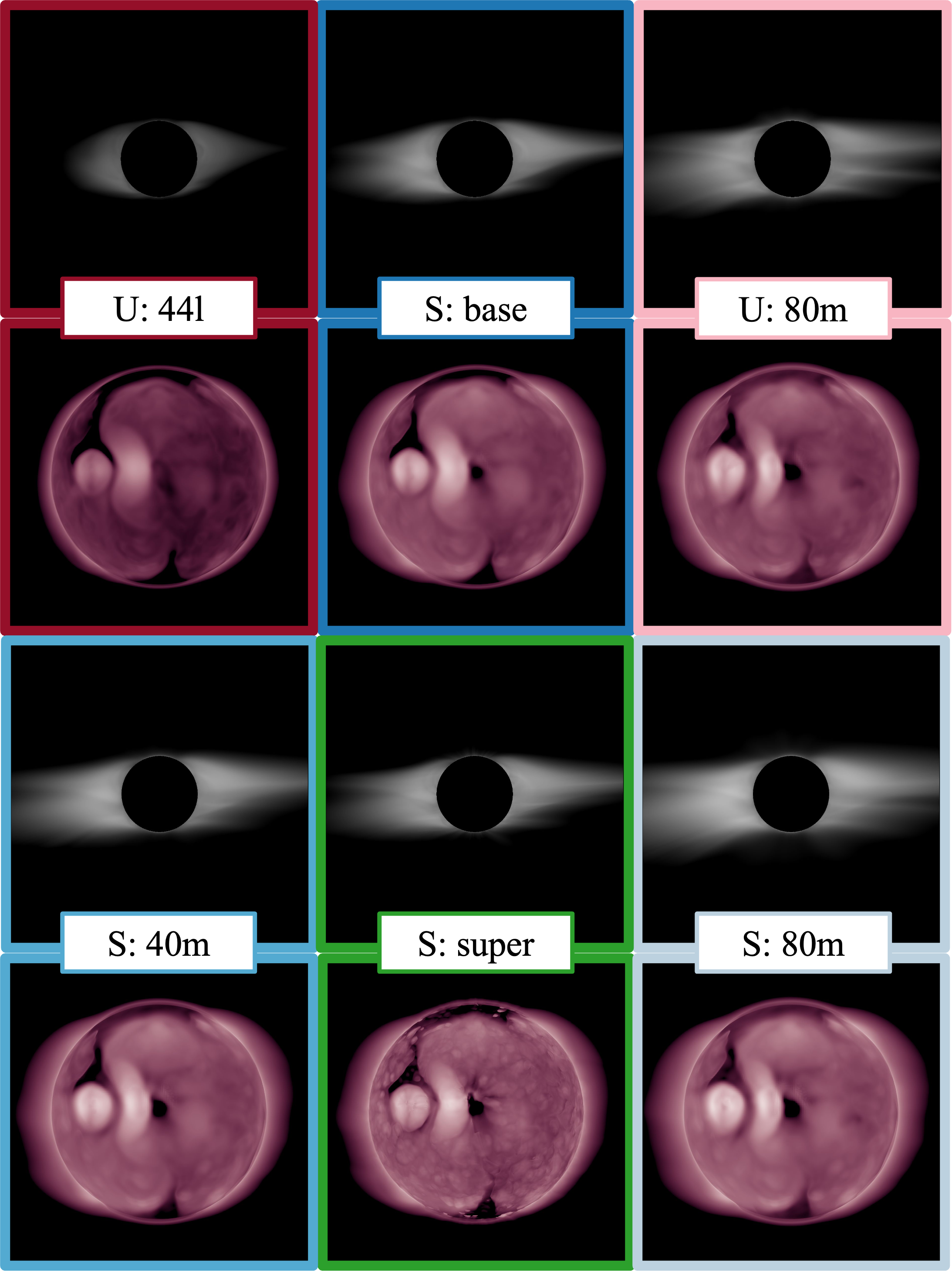}
        \caption{Representative forward modeled observables from Sets S and U. Rows 1 and 3 show radially filtered polarized brightness images, rows 2 and 4 show AIA 211 \r{A} images.  We include simulations with the lowest (U: 44l, in dark red, top left) and highest (S: 80m, lightest blue, bottom right) amount of Poynting flux in our experiment set. We further include S: base (top middle, darkest blue) for comparison to baseline parameters. U: 80m (pink, top right) and S: 40m (middle blue, bottom left) approximately have the same amount of net Poynting flux as the S: super simulation (green, bottom middle). Note that only the severely underheated U: 44l is demonstrably different.}
        \label{fig:context}
\end{figure*}

\resubi{We first examine morphology of coronal structures in the forward modeled images (see $\S$ \ref{sec:methods}). In Figure \ref{fig:context}, we present forward modeled polarized brightness and AIA 211 \r{A} images for a representative subset of our runs. We select the simulations with the most (S: 80m) and least  (U: 44l) amount of Poynting flux  and the reference ``base'' and super resolution runs. We also show our runs derived from the benchmark in \citet{Evans2026}: the U: 80m run and the S: 40m run. }

\resubi{There is large-scale qualitative similarity between the pB white-light images for S: base, S: 40m, and S: super resolution simulations. The U: 80m is more similar to the S: 80m simulation, with extra and enhanced streamers. For these highest z$_0$ values, both the east- and west-limb streamers expand. The U: 44l has depleted intensity and minimal structures present. This suggests that the overall amount of heating in the corona (as controlled by the value of z$_0$), is a key factor to determine large-scale, middle coronal morphology. }

\resubi{The synthetic 211 \r{A} images have differing intensities and morphologies, but similarities arise. For example, the eastern portion of the active region in both U: 80m and S: 80m have a similarly large spatial extent. These simulations also present an extended western portion of the southern polar coronal hole and a darker spot in the western quiet Sun region. These simulations have the same value of z$_0$ but different magnetic field scalings. In other words, while their morphologies are similar, they have visibly different overall levels of emission intensities and are thus observationally distinguishable.
Both decreased scaling and lesser z$_0$ combine to show significantly less emission in the U: 44l simulation, as well as an ill-defined equatorial coronal hole. We note that the S: super resolution EUV image has more smaller-scale structures than any other EUV image like the bright points in the coronal holes and more defined structures in the quiet Sun. This is a result of its improved resolution. In total, the simulations (excepting U: 44l) share the same large scale features; however there are key differences in morphologies and intensities. This indicates that key information about the subtleties of coronal hole formation, active region extent, and quiet Sun intensities differ between simulations.}

\begin{figure*}[hbt!] 
    \centering
    \includegraphics[scale=0.8]{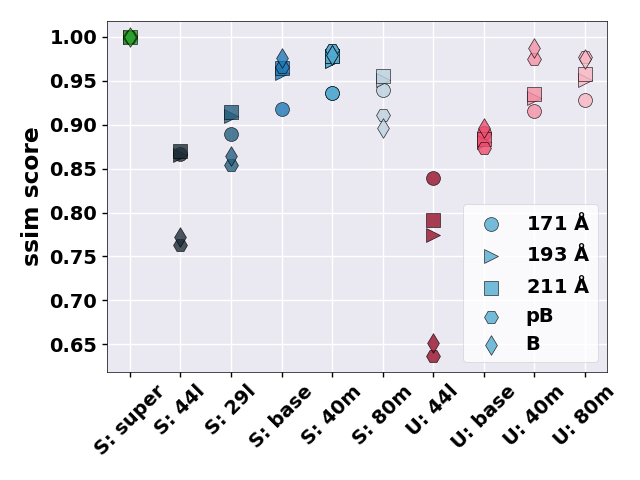}

    \caption{The structural similarity index (SSIM) score for each simulation as compared to the S: super run. We calculate for three EUV wavelengths measured by SDO/AIA (171, 193, and 211 \r{A} with circles, triangles, and squares respectively) and two white-light measurements: polarization brightness (pB, hexagon) and \resubi{total} brightness (B, diamond). Note the highest scores overall for S: 40m, improving on the white-light morphology in particular as compared to the also high-scoring U: 80m. }
    \label{fig:SSIM}
\end{figure*}

\resubi{We indicate the SSIM score for both the synthetic polarized brightness (pB) and \resubi{total} brightness (B) white-light as well as the 171, 193, and 211 \r{A} EUV images in Figure \ref{fig:SSIM}. As stated in $\S$ \ref{sec:ssim}, this metric ensures meaningful comparison among identically created images of different resolution simulations. We use the S: super simulation (green) as our reference solution. By definition for identical images, the SSIM score for comparing S: super to S: super is 1. Across both EUV and white-light images, there is general large-scale agreement, with the exception of dramatic underheating. Both the S: 44l and U: 44l have demonstrably lower SSIM scores for all observables in question. All of S: base, S: 40m, U: 40m, and U: 80m perform well, but the simulation that performs best overall for all observables is the S: 40m. While certain SSIM scores are similar among simulations, this illustrates how using multiple observables can help distinguish simulations from one another.}

\subsection{Magnetic Effects: Open Flux} \label{sec:magnetic}

First, we examine changes to the coronal magnetic field that result from our selected parameter space (see $\S$ \ref{sec:methods}). We focus primarily on the changes in the large-scale morphology and amount of open flux. 

\resubi{We calculate the integrated value of the open magnetic \resubi{flux (given} in $10^{22}$ Mx\resubi{)} in \resubi{Table \ref{tab:OF}}. Using the MapFL\footnote{https://github.com/predsci/MapFL} tool, we map field lines from the inner boundary surface of the model. A point at the inner surface contributes to the open flux if its field line connects to outer boundary surface (30 R$_\odot$). }

\begin{deluxetable*}{cccccc}[hbt!]
\label{tab:OF}
\tablecolumns{6}
\tablewidth{0pt}
\tablecaption{Open Flux Summary} 
\tablehead{\colhead{Set S, scaled} & 
\colhead{\shortstack{44\% less \\ (S: 44l)}} & 
\colhead{\shortstack{29\% less \\ (S: 29l)}} & 
\colhead{\shortstack{base/super\\ (S: base/S: super)}} & 
\colhead{\shortstack{40\% more\\ (S: 40m)}} & 
\colhead{\shortstack{80\% more \\ (S: 80m)}}}
\startdata
Open Flux (10$^{22}$ Mx) & 4.6019 & 4.8857 & 5.2647/5.6116 & 5.6097 & 5.8554 \\
Open Flux Fraction  & 0.40 & 0.43 &  0.46/0.27 & 0.49 & 0.51 \\
\hline 
\hline 
  & 44\% less  &  -  &  base   & 40\% more    &  80\% more   \\
 Set U, unscaled & (U: 44l)  &  -  &  (U: base)  & (U: 40m)   &   (U: 80m)   \\
\hline
Open Flux (10$^{22}$ Mx)   &  3.4869  &  -  &  3.8890  &  4.1547 &  4.2837 \\
Open Flux Fraction  &  0.43  & - & 0.48  & 0.51  & 0.53
\enddata
\end{deluxetable*}

\resubi{Increasing z$_0$ increases the open flux within each set. The magnitude of open flux is greater for Set S than Set U. As Set U has a weaker magnitude for the photospheric magnetic field, it is unsurprising that its open flux is smaller. Even if the morphology of the open flux map (see below for further discussion) captures similar areas, the lesser magnetic field translates to a smaller magnitude of open flux. }

\resubi{We also include the open flux fraction, which is the ratio between open flux and the total unsigned flux from each simulation's input boundary condition. This allows us to compare the fractional amount of open flux between set S and set U because the magnetogram scaling factor divides out in this ratio. Here, we see that the set U simulations are slightly more open than the comparable simulations in set S. This matches the qualitatively more open features in Figure \ref{fig:context} between U: 80m and S: 80m in both white light and EUV synthetic observations. }

\resubi{However, the open flux fraction can change significantly as a function of the $B_r$ boundary condition and model resolution. For example, the S: super simulation, as discussed in \citet{Evans2026}, has substantially higher unsigned magnetic flux than the lower resolution cases by virtue of its resolution, but a similar amount of open flux because the open areas map to flux concentrations with stronger magnetic fields distributed over smaller areas. In other words, this fraction is best used for inter-comparison between set S and set U and not across resolution.}

\resubi{In total, there is excellent correspondence between the open fluxes of S: 40m and S: super. However, this near numerical match represents an increase of less than 10\% between the S: base and S: 40m. The amount of open flux for U: 80m does not match. This provides initial evidence that increasing heating by tuning z$_0$ provides similar results to including smaller-scale photospheric flux \citep[as in][]{Evans2026} without overly opening the corona. The net Poynting flux does not perfectly match between these two simulations (see Table \ref{tab:sum}), but is fairly similar, and closer than S: base. While the same is true for U: 80m -- which also has reasonable SSIM scores -- we prefer S: 40m  because of its match to S: super for the open flux.}

\begin{figure*}[hbt!] 
   \centering

    \includegraphics[scale=0.36]{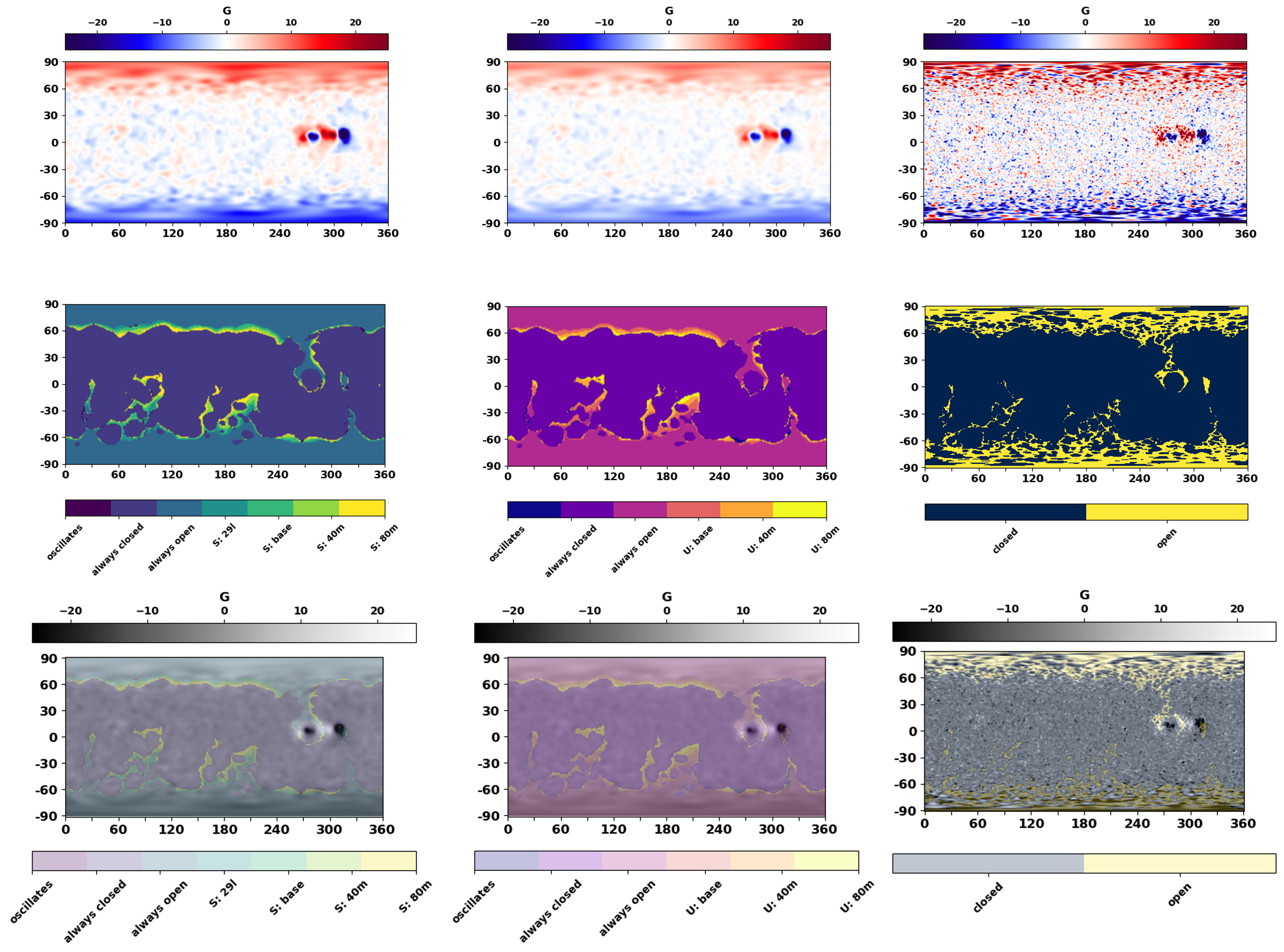}

    \caption {\resubi{Top row: }Photospheric inner radial magnetic boundary condition (across Carrington longitude and latitude in Gauss) \resubi{for} Set S (\resubi{leftmost)}, Set U (\resubi{middle}), and S: super (\resubi{rightmost). For Sets S and U, the background magnetic field is identical in morphology but differs in scaling. Middle row: } where the magnetic field is open (connecting to the outer boundary of the simulation) or closed (reconnecting to the inner boundary) \resubi{for Set S (leftmost) and Set U (middle)}. The colorbars indicate whether a given point remains closed throughout a simulation set (always closed), remains open (always open), or changes (oscillates). Otherwise, the subsequent labels specify for which simulation with increasing z$_0$ the point transitions to contributing to the open flux. The amount of open flux increases with z$_0$. \resubi{The rightmost panel in the middle row} shows the open (\resubi{yellow}) and closed (\resubi{blue}) flux for the S: super simulation. The morphology of the open flux is more structured given the simulation's enhanced resolution. \resubi{Bottom row: the appropriate photospheric inner radial magnetic boundary condition, overlaid with where the simulation (right) set (left and center) becomes open. Note the increasingly open morphology near the active region.}}
    \label{fig:OF_bound}
\end{figure*}

In Figure \ref{fig:OF_bound}, we examine the morphology of the quantity of open flux (see Table \resubi{\ref{tab:OF}}). \resubi{In the top row, we plot the photospheric inner boundary condition for all new simulations in this work for set S (left), set U (middle), and the preexisting set S: super (right). Blue (red) indicates a negative (positive) magnetic field in Gauss. In the middle row, we focus on open field morphology. For the middle left and middle right panels, we track how the open field morphology evolves with increasing values of z$_0$. Within a simulation set, if a mesh point always connects back to the photosphere, we label it ``always closed.'' If this mesh point remains part of the open field for all of the simulations in a set, we indicate it as ``always open.'' When a mesh point transitions from being closed to open field and remains so for all other increasing values of z$_0$, we indicate at which simulation it became open (e.g. S: 40m in lime green in the middle left). If a given mesh point changes between open and closed within the simulation set, we designate it as ``oscillates.'' For the rightmost middle plot, we show the open (yellow) and closed (blue) connectivity for the Set S: super resolution simulation. } 
\resubi{In the bottom row, we show which parts of the photospheric boundary condition (plotted from negative (black) to positive (white) magnetic field, Gauss) contribute to the open flux, combining the previous two rows.}

For both Sets S and U, as z$_0$ increases, the resulting heating increase also increases the open flux. For both simulation sets, the same regions in the synoptic map open with additional Poynting flux from higher values of z$_0$. These include a ring around half of the active region and various equatorial open patches of flux connecting to the open polar field. The extent of the open polar field additionally grows with increasing z$_0$.  

One possible solution to the open flux problem includes bright emission obscuring observationally-derived open flux. Observationally, the proxy for open fields is the EUV definition of the dark plasma of a coronal hole \citep[see discussion in][]{Cranmer2009}, though the exact definition and extent of open flux within a coronal hole remains an open research area \citep{Linker2021, Reiss2021, Arge2024, Leisner2026}. Regardless, active region EUV emission can obscure plasma that would otherwise be detected as a coronal hole contributing to the open field \citep[see similar work in][]{Wang2017, Caplan2023}. Moreover, from a modeling perspective, the presence and extent of an active region can influence coronal hole boundaries and change the amount of open flux \citep{Riley2006}. Since the opening up of flux near the active region occurs across both Sets S and U, this presents evidence that it is not just the strength of the magnetic field that determines the open flux but also the heating of the corona.

Accurate modeling of coronal holes is key for accurate solar wind prediction. The polar region morphology of open and closed flux \resubi{for Set S: super} does not match any of\resubi{ those in Sets S or U}. Although the net signed flux of the polar regions remains the same, the super resolution boundary condition resolves more patches of distinct bipolar magnetic flux elements beyond $\pm 60 \degr$ latitude, whereas the lower resolution cannot. Consequently, there are additional low-lying small-scale closed connectivities that cannot be represented. However, the extent of how the polar regions connect to the equatorial coronal holes are somewhat captured in the augmented case of S: 40m. By $0\degr$ latitude, $90\degr$ longitude, there is a patch of open flux in the super resolution simulation that the S: 40m recreates without opening beyond this patch, as the S: 80m simulation does. The U: 80m case in also has too large of a patch of open flux in this area.

Lastly, the morphology near the active region and the quantity of open flux in Table \resubi{\ref{tab:OF}} for S: 40m best matches S: super. This suggests that the derived quantity of 40m heating from \citet{Evans2026} is a viable ``subgrid'' heating correction. However, this is just one quantity that matches and it is a subtle increase ($< 10\%$). So while our modification of z$_0$ provides the necessary Poynting flux to emulate the open flux of the super simulation, we further investigate how our experiment modifies the hydrodynamics of the resulting simulation.

\subsection{Thermodynamic Effects} \label{sec:therm}
\subsubsection{Coronal Heating}

In this section, we examine how a uniform adjustment to WTD heating influences the spatial scales of heating throughout the corona. As described in \S\ref{sec:MAS}, we can specify the outward WTD wave amplitude, z$_0$, at the inner boundary. We examine heights from 1.01 to 1.51 R$_\odot$ to focus our analysis where most of the heating occurs in the low corona. This follows the same computational process as \citet{Evans2026}, including the same domain. However in this case, we subtract the contribution from the two spherically symmetric exponential heating terms that exist in every model to isolate the WTD heating contribution. As in \citet{Mikic2018}, these terms ensure a minimum floor to heating in the transition region and low corona as well as in open-field regions. These amount to \resubi{$3.57\times 10^{27} $ erg/s and 3.09$\times 10^{26} $ erg/s}, respectively. For reference, the two subtracted terms are specified volumetrically via exponentials of the form $H = H_0 \ \mathrm{exp}^{-(r-R_\odot)/\lambda}$. For the term targeting the transition region and low corona, $H_0 = 2.68 \! \times \!10^{-5}$ \resubi{erg/s/cm$^{3}$ }and $\lambda = 20.88$ Mm. For the open-field heating, $H_0 = 1.46 \! \times \! 10^{-8}$ \resubi{erg/s/cm$^{3}$} and $\lambda = 696$ Mm.

\begin{figure*}[hbt!] 

    \centering
    \includegraphics[scale=0.6]{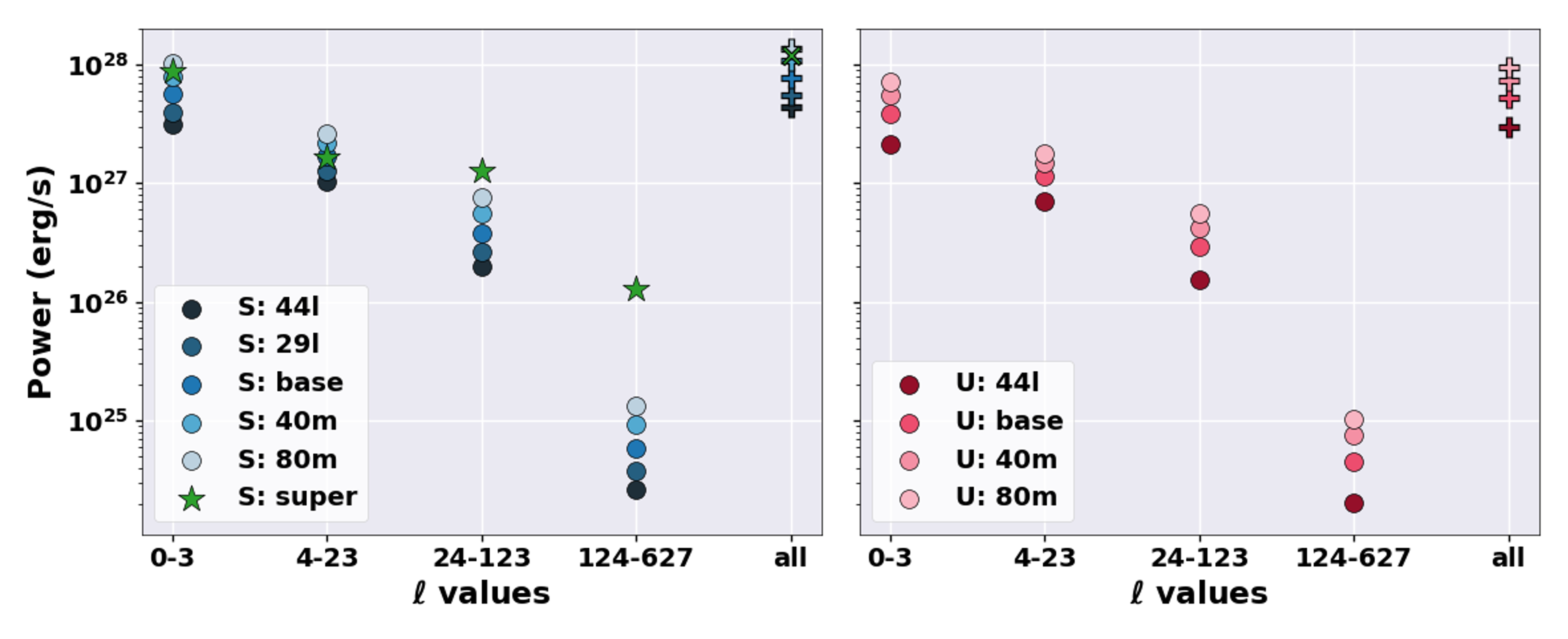}

    \caption{ \resubi{Integrated, volumetric }WTD heating of the corona \resubi{(ranging logarithmically)} spanning from 1.01 to 1.51 R$_\odot$, separated by logarithmically spaced bins of increasingly small spatial scales. The rightmost markers on each plot gives the total WTD heating for all spatial scales; on the left are the values for Set S (blues), including super (green, stars and x markers), and the right has the values for Set U (pinks). \resubi{By design,} Set S has more heating than Set U\resubi{; however,} and that increasing z$_0$ leads to increased volumetric heating within each set.}
    \label{fig:WTD_vol}
\end{figure*}

\begin{deluxetable*}{cccccc}[hbt!]
 \tablecolumns{6}
\tablewidth{0pt}
\tablecaption{Total Volumetric WTD Heating Ratios, 1.01 R$_\odot$ to 1.51 R$_\odot$ \label{tab:vol_heating}}
\tablehead{\colhead{} & \colhead{$\ell = [0, 3]$} & \colhead{$\ell = [4, 23]$} & \colhead{$\ell = [24, 123]$} & \colhead{$\ell = [124, 627]$} & \colhead{$\ell = [0, 627]$}}
 \startdata
 Set S & \quad & \quad & ratio & \quad & \quad \\
\hline
S: 44l/S: base & 0.55 & 0.62 & 0.52 & 0.45 & 0.57 \\
S: 29l/S: base & 0.71  & 0.76  & 0.69 & 0.64  & 0.72 \\
S: 40m/S: base & 1.42 & 1.29 & 1.48  & 1.58 & 1.39 \\
S: 80m/S: base & 1.83  & 1.54  & 1.99 & 2.26  & 1.78 \\
S: super/S: base & 1.56 &  0.97 &  3.36 & 21.65 &  1.53 \\
\hline 
\hline 
Set U & \quad & \quad & ratio & \quad & \quad \\
\hline
U: 44l/U: base & 0.55 & 0.61 & 0.53 & 0.45 & 0.56 \\
U: 40m/U: base & 1.42 & 1.30 & 1.45  & 1.68  & 1.40 \\
U: 80m/U: base & 1.85  & 1.56  & 1.92 & 2.30 & 1.80 \\
\enddata
\end{deluxetable*}

The total amount of WTD volumetric heating (erg/s) in the low corona as well as the fractional contributions separated by spatial scale are provided in Figure \ref{fig:WTD_vol}. Within both Sets S and U, more heating occurs on larger spatial scales and less so at the smaller spatial scales. Using SHD and Parseval's theorem, we separate the total amount of WTD heat into logarithmically spaced bins of $\ell =[0,3], [4, 23], [24, 123], \ \mathrm{and} \ [124, 627]$, where the first bin represents the largest spatial scales and the last bin corresponds to the smallest spatial scales. We indicate the total volumetric WTD heating ($\ell = [0,627]$) in the rightmost points (plus signs or x markers). Across all spatial scales and within both sets, the heating is ordered, where simulations with higher z$_0$ having more heat than those with lower z$_0$ values. 

We give the ratio to each set's base simulation in Table \ref{tab:vol_heating}. In isolating the WTD heating, comparing each simulation to its set's base simulation demonstrates how changing z$_0$ changes spatial scales within simulations with the same resolution. As in \citet{Evans2026}, the ratio in the $\ell = [0,3]$ bin approximately equals the ratio of the total ($\ell = [0,627]$) bin for all simulations. The ratios across Sets S and U -- e.g. S: 44l/S: base and U: 44l / U: base -- remain numerically similar. This indicates that despite the net Poynting flux being determined by a non-linear system of the coupled MHD and WTD  equations, the initial strength of the magnetic field does not strongly influence the relative heating ratios. 

However, for each of the cases where we decrease the heating (S: 44l, S: 29l, and U: 44l), the ratio of the $\ell = [4, 23]$ bin changes less than the ratio of the $\ell = [0,3]$ bin.  For this ratio to remain closer to one indicates that it was less influenced by changing z$_0$ than the $\ell = [0,3]$ bin. As compared to the increasing heat simulations (40m and 80m for Sets S and U), the relative reduction of the lesser heating simulations is stronger than the relative increase for these ratios.  As shown in Figure 1 of \citet{Evans2026}, the active region influences this $\ell = [4, 23]$ bin. Recalling the fact that increasing Poynting flux increases the amount of open flux, this suggests that for the increasing heat simulations, the more open active region would contribute less relative WTD heating in the low corona. This is because closed-fields generally dissipate more of the outward WTD Poynting flux as heating than open flux-tubes, which reflect more efficiently and advect this flux back through the inner boundary or outwards through the solar wind.

Both U: 80m and S: 40m approximately match the \resubi{volumetric} heating of the S: super resolution simulation. This shows that a reasonable \resubi{heating }match can be obtained by either increasing the \resubi{value of z$_0$} or by altering both \resubi{z$_0$} and surface magnetic field scaling in tandem. \resubi{This is in contrast to the differences noted in morphology $\S$ \resubi{\ref{sec:morph}} and open flux $\S$ \ref{sec:magnetic}. To evaluate whether these simulations are functionally equivalent representations of the S: super resolution simulation at a lower resolution, it is necessary to consider other plasma parameters beyond heating, as we do next.}

\subsubsection{Electron Temperature, Electron Density, and Emissivity Measure} \label{sec:EM_avgs}

In this section, we examine how varying the heating in both simulation sets changes the LOS integrated plasma properties of different regions of the Sun. We seek to establish how heating directly affects observable quantities for frequently observed solar features. We compute the overall \resubii{column-integrated} emission measure (EM) and the EM-weighted average electron temperature and density for representative regions for an active region (AR), coronal hole (CH), and quiet Sun (QS).

Mathematically, the EM is given as an LOS integral over the electron density ($n_e$), squared:
\begin{equation}
    EM = \int_{LOS} n_e^2 \ ds.
\end{equation}
 For each of the electron temperature ($T_{e}$) and density ($n_{e}$), the EM-weighted average is given as:
\begin{equation}
    <\!\vartheta\!> = \frac{\int_{LOS} n_e^2 \ \vartheta \ ds}{EM},
\end{equation}
where $\vartheta$ indicates the quantity to be averaged, either log$_{10}(T_{e})$ or log$_{10}(n_{e})$. 

\begin{figure*}[hbt!] 

    \gridline{\fig{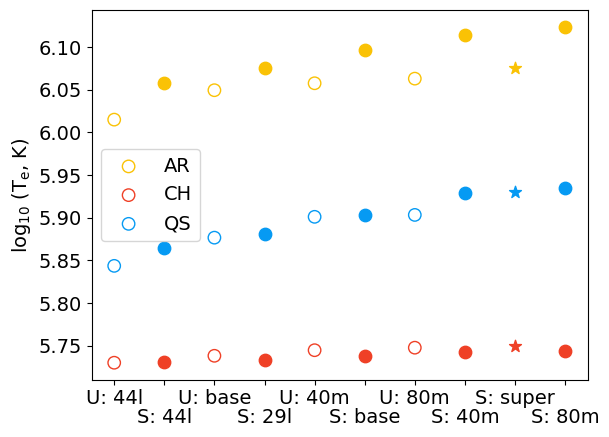}{0.49\textwidth}{(a)}
          \fig{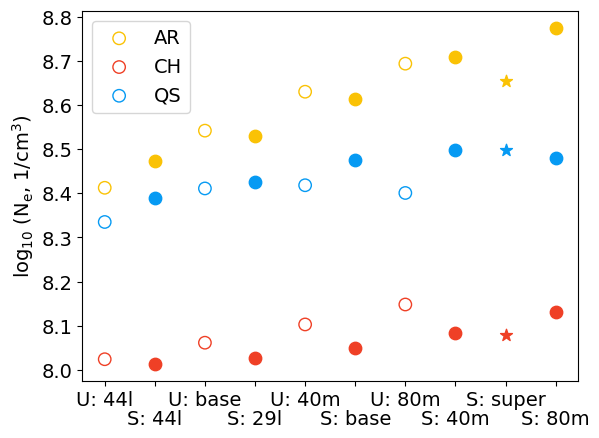}{0.49\textwidth}{(b)}}
    \gridline{\fig{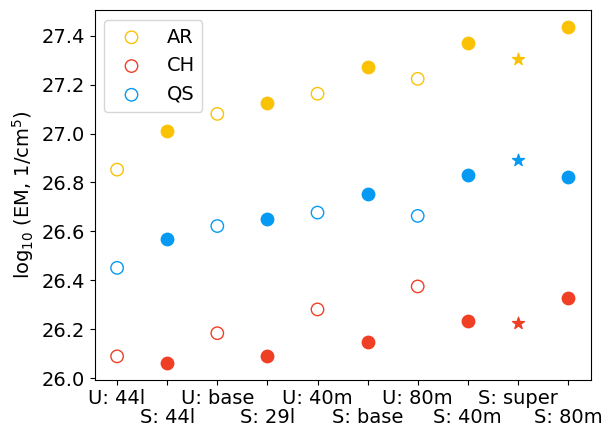}{0.49\textwidth}{(c)}
          \fig{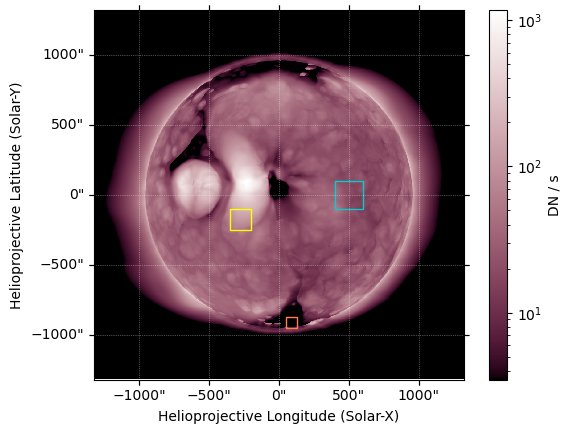}{0.49\textwidth}{(d)}}
   
    \caption{EM-weighted averages for the log$_{10}$ electron temperature (T$_\mathrm{e}$ in \resubi{K}, panel (a)) and electron density (N$_\mathrm{e}$ in \resubi{1/cm$^{3}$} panel (b)), alongside the log$_{10}$ of the \resubi{emission} measure (EM in \resubi{1/cm$^{5}$}, panel (c)) and context image of the S: super resolution. We select patches rooted in an active region (AR, yellow), coronal hole (CH, red), and quiet Sun (QS, blue) to compare quantities in panels (a-c) for each simulation, ordered by net Poynting flux (increasing from left to right). For Set U (unfilled), the density of the QS does not always increase with increasing Poynting flux. For Set S (filled), the same resolution simulations (i.e. excluding S: super) show direct scaling for most quantities. The S: 40m simulation mirrors the S: super simulation for many of these quantities, indicating its fidelity to the S: super simulation over U: 80m. }
    \label{fig:EM_avgs}
\end{figure*}

We examine the EM and EM-weighted averages not by separating by simulation set but by solely the magnitude of the net Poynting flux. In Figure \ref{fig:EM_avgs}, we present these averages, alongside our reference S: super simulation 211 \r{A} image indicating the fiducial AR (yellow), CH (red), and QS (blue) patches for our analysis. For each of the logarithm of the electron temperature (panel (a)), of the electron density (panel (b)), and of the EM (panel (c)), we show Set S as filled circles and Set U as unfilled circles.

For panels (a-c), the AR has a greater EM-weighted averaged quantity than the QS and both have larger values than the CH. All three panels also generally show, within each set, increasing EM-weighted quantities with increasing Poynting flux. The S: super simulation does not always follow this trend for all features (namely the AR). We further note disagreement for U: 80m\resubi{ for the QS patch for the electron density and the EM}.

The average log$_{10}$ of the electron density (panel (b)) varies somewhat more than the electron temperature (panel (a)). For the simulations with less Poynting flux, the distinction between the density of the QS and the AR patches lessens. Without sufficient heating, the AR will have lower densities, making it more similar to the QS. We point out the slight dark spots present in Figure \ref{fig:context} for both set's 80m runs. This is an indication of lower densities and tendency for a more open corona near this fiducial QS patch with sufficient heating. Panel (c), showing the EM variations, displays similar ordered trends within Set S and Set U outside of the 80m runs within the QS. At the far end of Set U (U: 80m), the difference in EM between the QS and CH for is smaller than any other simulation. This is consistent with Figure \ref{fig:context} for U: 80m, where the emission in the southern polar CH is brighter than other simulations, while a portion of the QS patch has become open (Figure \ref{fig:OF_bound}).

These results are consistent with relatively larger plasma $\beta$, achieved by lowering the magnetic field while increasing the heating. This simultaneously causes the CH feature to brighten relative to other simulations (higher density, still low $\beta$) but the QS feature levels off as the hydrodynamic forces increasingly dominate the magnetic field and cause the QS field lines to lengthen and ultimately open, reducing the relative EM increase that would otherwise be expected due to increasing heating. We further analyze the differences in plasma $\beta$ in $\S$ \ref{sec:beta}.

As for replicating the S: super run at a lower resolution, we focus on the simulations with the most similar net Poynting flux. In panel (a), the CH and QS electron temperatures are best matched by the S: 40m and S: 80m runs. However, both overestimate the temperature of the AR. Both the U: 40m and U: 80m runs slightly underestimate the temperature of the QS, but \resubi{present} reasonable matches for both the AR and CH. As seen in Figure \ref{fig:context}, the AR extent is narrower for both U: 80m and S: 80m simulations than for S: super. Thus the \resubi{loop} morphology of these ARs differs, leading to differences in AR temperatures.

In panel (b), the S: base run has similar values (unsurprising considering the identical parameterization and only differing resolution), but the S: 40m simulation best replicates the electron density of the S: super simulation for all three patches. This is likely a result of the closer net Poynting flux match between S: super and S: 40m (see Table \ref{tab:sum}). The AR density of S: 40m exceeds that of S: super, which could reflect that changing z$_0$ at a lower resolution inherently puts more heat on larger spatial scales (as shown in Table \ref{tab:vol_heating}).

Lastly, the EM of S: 40m shows the best agreement with the S: super resolution in the QS and CH regions. The next closest simulation with respect to net Poynting flux is the U: 80m simulation, which has far too much emission in the CH, as discussed above. This suggests that, at least for solar minimum conditions, a uniform enhancement of heating via z$_0$ is sufficient to replicate the super resolution model. However, the agreement with the AR is least precise. This could be due to the lower resolution of S: 40m compared to S: super. A lower resolution grid is less able to resolve distinct thermal non-equilibrium (TNE) cycles along adjacent loops, which can lead to differences in the mixture of hot and cool plasma within a given active region \citep{Mok2016, Downs2016}.

\subsection{Combined Magnetothermal Effects: Plasma $\beta$} \label{sec:MHD_effects}
\label{sec:beta}

\begin{figure*}[hbt!] 
   \centering

   \includegraphics[scale=0.42]{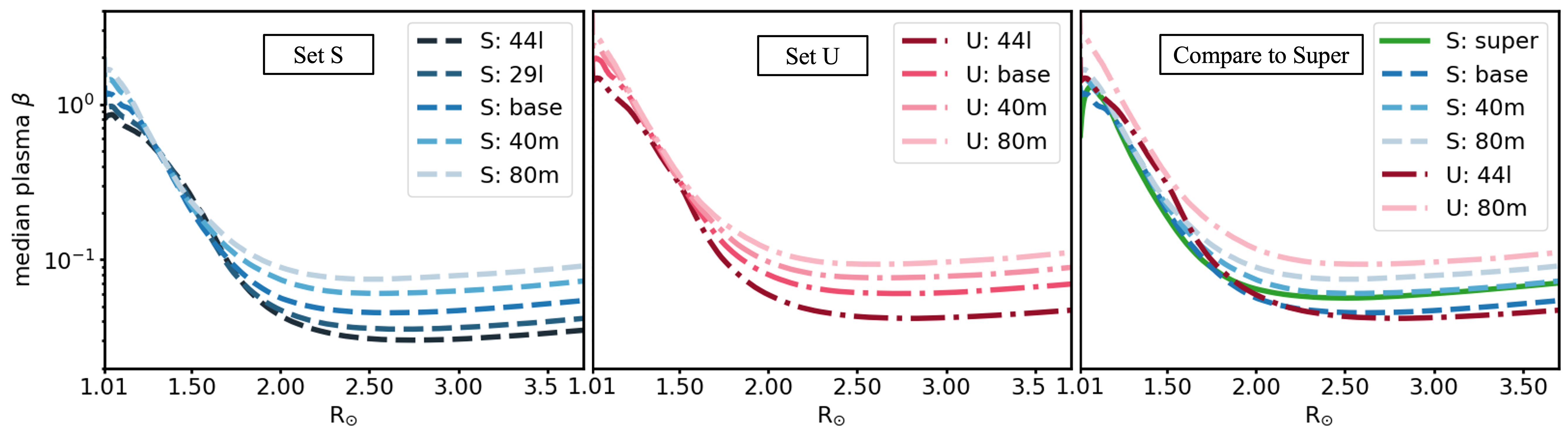}
   
    \caption{The area-weighted median plasma $\beta$ for the same resolution of Set S (\resubi{leftmost}), Set U (middle), and simulations of interest as indicated in Figure \ref{fig:context} (\resubi{rightmost}) with respect to radial height (spanning 1.01 - 3.7 R$_\odot$). Overall, decreasing Poynting flux decreases the values of plasma $\beta$ within each set. As seen in the rightmost panel, the decreased strength of the magnetic field of Set U raises the values of plasma $\beta$. Notice the agreement between the evolution of plasma $\beta$ for S: 40m and S: super.}
    \label{fig:beta}
\end{figure*}

As discussed in previous sections, the overall morphology and relative openness of the corona \resubi{change} depending on the amount of heating and scaling of the magnetic field. To understand how these changes are mediated by the interplay between hydrodynamic and magnetic forces, we examine plasma $\beta$ in the volume of each simulation. In Figure~\ref{fig:beta}, we show the area-weighted median plasma $\beta$ for Set S (left, blues), Set U (center, pinks), and their comparison to S: super (in the rightmost panel). We plot this with respect to radial coordinate (in solar radii, R$_\odot$) from 1.01 to 3.7 R$_\odot$. As each simulation set has the same initial boundary magnetic field, it is expected that plasma $\beta$ should increase with increasing heating (z$_0$) and subsequent density increases. Between Sets S and U, the latter has higher values of plasma $\beta$ because the magnitude of the magnetic field decreases. Within both sets, the simulations with more heating have a higher value of plasma $\beta$ as a result of increasing density (generally seen in panel (b) of Figure \ref{fig:EM_avgs}). 

Other evidence of a larger \resubi{value} of plasma $\beta$ \resubi{between sets} includes morphological differences in the largest-scale structures. In the K-corona images in Figure \ref{fig:context}, both the U: 80m and S: 80m have more \resubi{streamers} that are more open than any of the four other simulations shown. Within the 211 \r{A} images, the equatorial CH has greatly diminished in the lowest plasma $\beta$ simulation and grows in size with increasing values of plasma $\beta$. These are large-scale, observable differences.

Plasma $\beta$ mediates the amount of open flux. In Table \ref{tab:sum}, we calculate open flux for each simulation within both sets. The process to calculate open flux depends on field-line mapping to derive the area of open flux on an inner boundary condition then multiplying that area by the field of the inner boundary condition. To quantitatively compare the amount of open flux in the context of plasma $\beta$, we must account for the scaling difference between Set S and Set U. Since these are otherwise identical inner boundary conditions, we therefore scale the values in Table \ref{tab:sum} for Set U by 1.4 to compare to Set S. This gives the U: 44l simulation an adjusted open flux of $4.8817 \times 10^{22}$ Mx and U: 80m simulation an adjusted open flux of $5.9972 \times 10^{22}$ Mx. These adjusted open fluxes are the least and greatest values of open flux for the simulations highlighted in the right panel of Figure \ref{fig:beta}. This matches the ordering where U: 80m has the largest plasma $\beta$ and U: 44l has the smallest. 

The simulations of Set U are relatively more open than those of Set S. The critical comparison is between the open fluxes of S: 80m ($5.8554 \times 10^{22}$ Mx) and U: 80m (adjusted to $5.9972 \times 10^{22}$ Mx). The U: 80m simulation is more open than S: 80m. This indicates that the open flux does not drop by a factor of 1.4 for Set U; rather, the Set U open flux occupies a larger area than that of Set S.  \resubi{This is also seen in the open flux fraction in Table \ref{tab:OF}.} Beyond this case, Set U has generally higher median plasma $\beta$ in the middle corona than Set S, making plasma $\beta$ a key indicator for open flux.

The mechanism in which we increase Poynting flux changes the median value of plasma $\beta$, meaning both heating and magnetic field scaling cannot be free parameters as they drive coronal morphology and open flux. The rightmost panel indicates that plasma $\beta$ is key for accurate observables. For U: 80m, the median plasma $\beta$ values exceed the values of all other simulations for the entire radial domain of interest. Its lack of scaling may explain why the U: 80m simulation compares somewhat favorably to S: super in thermodynamic applications (as seen in $\S$ \ref{sec:therm}), but otherwise has a poorer match for magnetic quantities ($\S$ \ref{sec:magnetic}) and observables ($\S$ \ref{sec:EM_avgs}). The S: 40m well-matches the median value of plasma $\beta$ of S: super, as do the heating, open flux, and observables. This suggests that the S: 40m replicates what is plausible to simulate at a lower resolution. 

\section{Conclusions} \label{sec:conclusions}

In this study, we explored how changing the overall amount of heating and/or magnetic field strength influences properties of the global solar corona. We both (i) modified the surface Poynting flux of Alfv\'en waves in a wave-turbulence-driven model of coronal heating and (ii) scaled the magnetic field strength of the surface boundary condition. All new simulations presented in this work otherwise have identical meshes and parameter spaces. We \resubi{first} examined the effects on key plasma properties and observables of the corona.

Second, we compared our modest resolution runs to an extremely high-resolution simulation \citep{Evans2026} that incorporated high-resolution surface magnetic fields (and a larger unsigned flux). As any model can be fine-tuned for a particular outcome, we compared multiple hydrodynamic, magnetic, and magnetohydrodynamic facets of our simulations to this reference super-resolution simulation to investigate which simulation best replicates it. This is a key part of validating  numerical models and testing model optimizations.

We generated forward models to complement analysis of our MHD state variables. We utilized an image comparison metric (the structural similarity index, SSIM), traced field lines to calculate the open flux, \resubi{and} performed spherical harmonic decomposition to examine volumetric heating across spatial scales. We also calculated the emission measure and \resubi{emission-measure}-weighted averages for representative morphological regions of the EUV corona, and computed the area-weighted median plasma $\beta$ in the low and middle corona.  
We summarize our findings as follows.

1. Increasing the input heating of the simulation opens up the corona and increases the overall heating. This results in a larger open flux.

2. The scaling of the magnetic field is not a free parameter for MHD; its relative strength is key to determining the morphology of the corona.

3. As expected, plasma $\beta$ increases for simulations with lower magnetic field and also with increasing heating. A larger median value of $\beta$ in the middle corona strongly correlates to a more open morphology in the K-corona and a greater relative amount of open magnetic flux.

4. We developed a set of metrics to determine how well a simulation of a lower resolution emulates a higher resolution. We find that simply increasing the overall heating by $\sim40\%$ best emulates the magnetic, thermodynamic, and combined magnetothermal properties of the reference super-resolution simulation. This includes key global observables like EUV and white-light images. \resubi{Decreasing the scaling of the magnetic field by $\sim40\%$ and increasing the heating by $\sim80\%$ does not replicate the super-resolution simulation as well across all metrics.}

It is important to note that we targeted the solar minimum conditions of 2019 for this study, primarily to minimize the overall complexity of the global magnetic field, and highlight the overall changes to the large-scale streamer structure. It is likely that some of the relative changes in coronal properties may differ during maximum conditions, when a larger fraction of coronal heating and open magnetic flux can be associated with active regions. That said, we expect the general trends of an increasingly more open corona with increased heating and/or reduced magnetic field strength to hold, and we will explore this in a future study. \resubi{We will also investigate how localized changes to the net polar flux, which is challenging to estimate from current observations, might influence the resulting field morphologies and thus heating and plasma properties.}

Ultimately, this study illustrates how the scaling of the magnetic field and the heating of the corona are fundamentally intertwined with the open flux, morphology, and plasma properties of the global solar corona. Given that photospheric magnetic field measurements may underestimate the absolute value of magnetic flux across the solar disk \citep{Petrie2025, Milic2024, Centeno2023}, and the open flux problem remains unsolved \citep{Linker2017, Arge2024}, it is important to explore the combined consequences of magnetic field scaling when studying the coronal heating and solar wind acceleration problems from a global modeling perspective. 
\begin{acknowledgments}

This work was supported by the University of Colorado Boulder George Ellery Hale Graduate Fellowship and the NASA Living With a Star program, in particular the Focused Science Topic: Towards a Quantitative Description of the Magnetic Origins of the Corona and Inner Heliosphere (grant 80NSSC22K1021). It was also supported by the NASA Heliophysics Supporting Research program (grant 80NSSC18K1129). Resources supporting this work were provided by the NASA High-End Computing (HEC) Program through the NASA Advanced Supercomputing (NAS) Division at Ames Research Center. This work additionally used Expanse at the San Diego Supercomputing Center (SDSC) at UC San Diego through allocation TG-MCA03S014 from the Advanced Cyberinfrastructure Coordination Ecosystem: Services \& Support (ACCESS) program, which is supported by National Science Foundation grants \#2138259, \#2138286, \#2138307, \#2137603, and \#2138296. This research used version 7.0.1 of the SunPy open source software package \citep{SunPy2020}. We thank Maria Kazachenko for interesting discussions and support of this work.
\end{acknowledgments}

\section{Appendix A: Forward Modeling} \label{sec:appendix}

We expand on the heuristic treatment for strong coronal lines mentioned in $\S$ \ref{sec:forwardmodel}. The same ion and transition can resonantly scatter at different locations of the atmosphere \citep[e.g.][]{SchrijverMcMullen2000}, leading to reduced or redistributed emission. For example, the bright footpoint or loop emission in an active region may be subject to additional scattering \citep{WoodRaymond2000}. However, accurate radiative transfer that accounts for both optically thin and optically thick absorption and re-emission (resonant scattering) is computationally prohibitive given the complexity and dimensionality of the problem. Instead, as discussed in \citet{Mok2016},  the intensity reduction can be emulated approximately by adding a small opacity that behaves like continuum absorption for the relevant ions of a given passband (e.g. Fe IX for 171 \r{A}, Fe XI for 211 \r{A}, and Fe XII for 193 \r{A}, per \citet{ODwyer2010}).

We accomplish this by uniformly reducing the cross-section for a given line that is computed from the atomic parameters of the transition by a factor of 100 and then using this cross section in Eq.~\ref{eq:emission}. The \resubi{resulting} values respectively are: Fe IX of $ 2.44 \times 10^{-15}$, Fe XI of $1.28 \times 10^{-17}$, Fe XII of $4.14 \times 10^{-16}$, and Fe XIV of $3.20 \times 10^{-16}$ cm$^2$. This is a functional way to include opacity induced by dense material. 

On a related point, the MAS code uses a technique to broaden the transition region in thermodynamic MHD calculations so that the energy transfer across it can be captured in 3D \citep{Lionello2009}. This involves modifying the thermal conduction coefficient below a certain cutoff temperature (in this case 400,000K). We similarly reduce the emissivity by \resubi{the same reduction factor} number in the forward modeling calculation to account for this broadened path-length.

\bibliography{paperII}{}

@ARTICLE{Downs2016,
       author = {{Downs}, Cooper and {Lionello}, Roberto and {Miki{\'c}}, Zoran and {Linker}, Jon A. and {Velli}, Marco},
        title = "{Closed-field Coronal Heating Driven by Wave Turbulence}",
      journal = {\apj},
         year = 2016,
        month = dec,
       volume = {832},
       number = {2},
          eid = {180},
        pages = {180},
          doi = {10.3847/0004-637X/832/2/180},
archivePrefix = {arXiv},
       eprint = {1610.02113},
 primaryClass = {astro-ph.SR},
       adsurl = {https://ui.adsabs.harvard.edu/abs/2016ApJ...832..180D}
}

@ARTICLE{Mikic2018,
       author = {{Miki{\'c}}, Zoran and {Downs}, Cooper and {Linker}, Jon A. and {Caplan}, Ronald M. and {Mackay}, Duncan H. and {Upton}, Lisa A. and {Riley}, Pete and {Lionello}, Roberto and {T{\"o}r{\"o}k}, Tibor and {Titov}, Viacheslav S. and {Wijaya}, Janvier and {Druckm{\"u}ller}, Miloslav and {Pasachoff}, Jay M. and {Carlos}, Wendy},
        title = "{Predicting the corona for the 21 August 2017 total solar eclipse}",
      journal = {Nature Astronomy},
         year = 2018,
        month = aug,
       volume = {2},
        pages = {913-921},
          doi = {10.1038/s41550-018-0562-5},
       adsurl = {https://ui.adsabs.harvard.edu/abs/2018NatAs...2..913M}
}

@ARTICLE{Wieczorek2018,
       author = {{Wieczorek}, Mark A. and {Meschede}, Matthias},
        title = "{SHTools: Tools for Working with Spherical Harmonics}",
      journal = {Geochemistry, Geophysics, Geosystems},
         year = 2018,
        month = aug,
       volume = {19},
       number = {8},
        pages = {2574-2592},
          doi = {10.1029/2018GC007529},
       adsurl = {https://ui.adsabs.harvard.edu/abs/2018GGG....19.2574W}
}

@ARTICLE{Mikic1999,
       author = {{Miki{\'c}}, Zoran and {Linker}, Jon A. and {Schnack}, Dalton D. and {Lionello}, Roberto and {Tarditi}, Alfonso},
        title = "{Magnetohydrodynamic modeling of the global solar corona}",
      journal = {Physics of Plasmas},
         year = 1999,
        month = may,
       volume = {6},
       number = {5},
        pages = {2217-2224},
          doi = {10.1063/1.873474},
       adsurl = {https://ui.adsabs.harvard.edu/abs/1999PhPl....6.2217M}
}

@ARTICLE{Lionello2009,
       author = {{Lionello}, Roberto and {Linker}, Jon A. and {Miki{\'c}}, Zoran},
        title = "{Multispectral Emission of the Sun During the First Whole Sun Month: Magnetohydrodynamic Simulations}",
      journal = {\apj},
         year = 2009,
        month = jan,
       volume = {690},
       number = {1},
        pages = {902-912},
          doi = {10.1088/0004-637X/690/1/902},
       adsurl = {https://ui.adsabs.harvard.edu/abs/2009ApJ...690..902L}
}

@ARTICLE{Lionello2013CME,
       author = {{Lionello}, Roberto and {Downs}, Cooper and {Linker}, Jon A. and {T{\"o}r{\"o}k}, Tibor and {Riley}, Pete and {Miki{\'c}}, Zoran},
        title = "{Magnetohydrodynamic Simulations of Interplanetary Coronal Mass Ejections}",
      journal = {\apj},
         year = 2013,
        month = nov,
       volume = {777},
       number = {1},
          eid = {76},
        pages = {76},
          doi = {10.1088/0004-637X/777/1/76},
       adsurl = {https://ui.adsabs.harvard.edu/abs/2013ApJ...777...76L}
}

@ARTICLE{Riley2011,
       author = {{Riley}, P. and {Lionello}, R. and {Linker}, J.~A. and {Mikic}, Z. and {Luhmann}, J. and {Wijaya}, J.},
        title = "{Global MHD Modeling of the Solar Corona and Inner Heliosphere for the Whole Heliosphere Interval}",
      journal = {\solphys},
         year = 2011,
        month = dec,
       volume = {274},
       number = {1-2},
        pages = {361-377},
          doi = {10.1007/s11207-010-9698-x},
       adsurl = {https://ui.adsabs.harvard.edu/abs/2011SoPh..274..361R}
}

@ARTICLE{Torok2018,
       author = {{T{\"o}r{\"o}k}, Tibor and {Downs}, Cooper and {Linker}, Jon A. and {Lionello}, R. and {Titov}, Viacheslav S. and {Miki{\'c}}, Zoran and {Riley}, Pete and {Caplan}, Ronald M. and {Wijaya}, Janvier},
        title = "{Sun-to-Earth MHD Simulation of the 2000 July 14 {\textquotedblleft}Bastille Day{\textquotedblright} Eruption}",
      journal = {\apj},
         year = 2018,
        month = mar,
       volume = {856},
       number = {1},
          eid = {75},
        pages = {75},
          doi = {10.3847/1538-4357/aab36d},
archivePrefix = {arXiv},
       eprint = {1801.05903},
 primaryClass = {astro-ph.SR},
       adsurl = {https://ui.adsabs.harvard.edu/abs/2018ApJ...856...75T}
}

@ARTICLE{DeForest2018,
       author = {{DeForest}, C.~E. and {Howard}, R.~A. and {Velli}, M. and {Viall}, N. and {Vourlidas}, A.},
        title = "{The Highly Structured Outer Solar Corona}",
      journal = {\apj},
         year = 2018,
        month = jul,
       volume = {862},
       number = {1},
          eid = {18},
        pages = {18},
          doi = {10.3847/1538-4357/aac8e3},
       adsurl = {https://ui.adsabs.harvard.edu/abs/2018ApJ...862...18D}
}

@ARTICLE{Boe2021,
       author = {{Boe}, Benjamin and {Habbal}, Shadia and {Downs}, Cooper and {Druckm{\"u}ller}, Miloslav},
        title = "{The Color and Brightness of the F-corona Inferred from the 2019 July 2 Total Solar Eclipse}",
      journal = {\apj},
         year = 2021,
        month = may,
       volume = {912},
       number = {1},
          eid = {44},
        pages = {44},
          doi = {10.3847/1538-4357/abea79},
archivePrefix = {arXiv},
       eprint = {2103.02113},
 primaryClass = {astro-ph.SR},
       adsurl = {https://ui.adsabs.harvard.edu/abs/2021ApJ...912...44B}
}

@ARTICLE{Boe2022,
       author = {{Boe}, Benjamin and {Habbal}, Shadia and {Downs}, Cooper and {Druckm{\"u}ller}, Miloslav},
        title = "{The Solar Minimum Eclipse of 2019 July 2. II. The First Absolute Brightness Measurements and MHD Model Predictions of Fe X, XI, and XIV out to 3.4 R $_{{\ensuremath{\odot}}}$}",
      journal = {\apj},
         year = 2022,
        month = aug,
       volume = {935},
       number = {2},
          eid = {173},
        pages = {173},
          doi = {10.3847/1538-4357/ac8101},
archivePrefix = {arXiv},
       eprint = {2206.10106},
 primaryClass = {astro-ph.SR},
       adsurl = {https://ui.adsabs.harvard.edu/abs/2022ApJ...935..173B}
}

@ARTICLE{Boe2023,
       author = {{Boe}, Benjamin and {Downs}, Cooper and {Habbal}, Shadia},
        title = "{The Solar Minimum Eclipse of 2019 July 2. III. Inferring the Coronal T $_{ e }$ with a Radiative Differential Emission Measure Inversion}",
      journal = {\apj},
         year = 2023,
        month = jul,
       volume = {951},
       number = {1},
          eid = {55},
        pages = {55},
          doi = {10.3847/1538-4357/acd10b},
archivePrefix = {arXiv},
       eprint = {2306.04826},
 primaryClass = {astro-ph.SR},
       adsurl = {https://ui.adsabs.harvard.edu/abs/2023ApJ...951...55B}
}

@ARTICLE{Parker1958,
       author = {{Parker}, E.~N.},
        title = "{Dynamics of the Interplanetary Gas and Magnetic Fields.}",
      journal = {\apj},
         year = 1958,
        month = nov,
       volume = {128},
        pages = {664},
          doi = {10.1086/146579},
       adsurl = {https://ui.adsabs.harvard.edu/abs/1958ApJ...128..664P}
}

@ARTICLE{Linker2017,
       author = {{Linker}, J.~A. and {Caplan}, R.~M. and {Downs}, C. and {Riley}, P. and {Mikic}, Z. and {Lionello}, R. and {Henney}, C.~J. and {Arge}, C.~N. and {Liu}, Y. and {Derosa}, M.~L. and {Yeates}, A. and {Owens}, M.~J.},
        title = "{The Open Flux Problem}",
      journal = {\apj},
         year = 2017,
        month = oct,
       volume = {848},
       number = {1},
          eid = {70},
        pages = {70},
          doi = {10.3847/1538-4357/aa8a70},
archivePrefix = {arXiv},
       eprint = {1708.02342},
 primaryClass = {astro-ph.SR},
       adsurl = {https://ui.adsabs.harvard.edu/abs/2017ApJ...848...70L}
}

@ARTICLE{Linker2021,
       author = {{Linker}, Jon A. and {Heinemann}, Stephan G. and {Temmer}, Manuela and {Owens}, Mathew J. and {Caplan}, Ronald M. and {Arge}, Charles N. and {Asvestari}, Eleanna and {Delouille}, Veronique and {Downs}, Cooper and {Hofmeister}, Stefan J. and {Jebaraj}, Immanuel C. and {Madjarska}, Maria S. and {Pinto}, Rui F. and {Pomoell}, Jens and {Samara}, Evangelia and {Scolini}, Camilla and {Vr{\v{s}}nak}, Bojan},
        title = "{Coronal Hole Detection and Open Magnetic Flux}",
      journal = {\apj},
         year = 2021,
        month = sep,
       volume = {918},
       number = {1},
          eid = {21},
        pages = {21},
          doi = {10.3847/1538-4357/ac090a},
archivePrefix = {arXiv},
       eprint = {2103.05837},
 primaryClass = {astro-ph.SR},
       adsurl = {https://ui.adsabs.harvard.edu/abs/2021ApJ...918...21L}
}

@ARTICLE{Lowder2017,
       author = {{Lowder}, Chris and {Qiu}, Jiong and {Leamon}, Robert},
        title = "{Coronal Holes and Open Magnetic Flux over Cycles 23 and 24}",
      journal = {\solphys},
         year = 2017,
        month = jan,
       volume = {292},
       number = {1},
          eid = {18},
        pages = {18},
          doi = {10.1007/s11207-016-1041-8},
archivePrefix = {arXiv},
       eprint = {1612.07595},
 primaryClass = {astro-ph.SR},
       adsurl = {https://ui.adsabs.harvard.edu/abs/2017SoPh..292...18L}
}

@ARTICLE{Wang2022,
       author = {{Wang}, Y. -M. and {Ulrich}, R.~K. and {Harvey}, J.~W.},
        title = "{Magnetograph Saturation and the Open Flux Problem}",
      journal = {\apj},
         year = 2022,
        month = feb,
       volume = {926},
       number = {2},
          eid = {113},
        pages = {113},
          doi = {10.3847/1538-4357/ac4491},
archivePrefix = {arXiv},
       eprint = {2112.09969},
 primaryClass = {astro-ph.SR},
       adsurl = {https://ui.adsabs.harvard.edu/abs/2022ApJ...926..113W}
}

@ARTICLE{Seaton2021,
       author = {{Seaton}, Daniel B. and {Hughes}, J. Marcus and {Tadikonda}, Sivakumara K. and {Caspi}, Amir and {DeForest}, Craig E. and {Krimchansky}, Alexander and {Hurlburt}, Neal E. and {Seguin}, Ralph and {Slater}, Gregory},
        title = "{The Sun's dynamic extended corona observed in extreme ultraviolet}",
      journal = {Nature Astronomy},
         year = 2021,
        month = oct,
       volume = {5},
        pages = {1029-1035},
          doi = {10.1038/s41550-021-01427-8},
archivePrefix = {arXiv},
       eprint = {2105.08028},
 primaryClass = {astro-ph.SR},
       adsurl = {https://ui.adsabs.harvard.edu/abs/2021NatAs...5.1029S}
}

@ARTICLE{Arge2024,
       author = {{Arge}, C. Nick and {Leisner}, Andrew and {Antiochos}, Spiro K. and {Wallace}, Samantha and {Henney}, Carl J.},
        title = "{Proposed Resolution to the Solar Open Magnetic Flux Problem}",
      journal = {\apj},
         year = 2024,
        month = apr,
       volume = {964},
       number = {2},
          eid = {115},
        pages = {115},
          doi = {10.3847/1538-4357/ad20e2},
archivePrefix = {arXiv},
       eprint = {2304.07649},
 primaryClass = {astro-ph.SR},
       adsurl = {https://ui.adsabs.harvard.edu/abs/2024ApJ...964..115A}
}

@ARTICLE{West2023,
       author = {{West}, Matthew J. and {Seaton}, Daniel B. and {Wexler}, David B. and {Raymond}, John C. and {Del Zanna}, Giulio and {Rivera}, Yeimy J. and {Kobelski}, Adam R. and {Chen}, Bin and {DeForest}, Craig and {Golub}, Leon and {Caspi}, Amir and {Gilly}, Chris R. and {Kooi}, Jason E. and {Meyer}, Karen A. and {Alterman}, Benjamin L. and {Alzate}, Nathalia and {Andretta}, Vincenzo and {Auch{\`e}re}, Fr{\'e}d{\'e}ric and {Banerjee}, Dipankar and {Berghmans}, David and {Chamberlin}, Phillip and {Chitta}, Lakshmi Pradeep and {Downs}, Cooper and {Giordano}, Silvio and {Harra}, Louise and {Higginson}, Aleida and {Howard}, Russell A. and {Kumar}, Pankaj and {Mason}, Emily and {Mason}, James P. and {Morton}, Richard J. and {Nykyri}, Katariina and {Patel}, Ritesh and {Rachmeler}, Laurel and {Reardon}, Kevin P. and {Reeves}, Katharine K. and {Savage}, Sabrina and {Thompson}, Barbara J. and {Van Kooten}, Samuel J. and {Viall}, Nicholeen M. and {Vourlidas}, Angelos and {Zhukov}, Andrei N.},
        title = "{Defining the Middle Corona}",
      journal = {\solphys},
         year = 2023,
        month = jun,
       volume = {298},
       number = {6},
          eid = {78},
        pages = {78},
          doi = {10.1007/s11207-023-02170-1},
archivePrefix = {arXiv},
       eprint = {2208.04485},
 primaryClass = {astro-ph.SR},
       adsurl = {https://ui.adsabs.harvard.edu/abs/2023SoPh..298...78W}
}

@ARTICLE{Lionello2023,
       author = {{Lionello}, Roberto and {Downs}, Cooper and {Mason}, Emily I. and {Linker}, Jon A. and {Caplan}, Ronald M. and {Riley}, Pete and {Titov}, Viacheslav S. and {DeRosa}, Marc L.},
        title = "{Global MHD Simulations of the Time-dependent Corona}",
      journal = {\apj},
         year = 2023,
        month = dec,
       volume = {959},
       number = {2},
          eid = {77},
        pages = {77},
          doi = {10.3847/1538-4357/ad00be},
archivePrefix = {arXiv},
       eprint = {2306.12551},
 primaryClass = {astro-ph.SR},
       adsurl = {https://ui.adsabs.harvard.edu/abs/2023ApJ...959...77L}
}

@ARTICLE{Riley2006,
       author = {{Riley}, Pete and {Linker}, J.~A. and {Miki{\'c}}, Z. and {Lionello}, R. and {Ledvina}, S.~A. and {Luhmann}, J.~G.},
        title = "{A Comparison between Global Solar Magnetohydrodynamic and Potential Field Source Surface Model Results}",
      journal = {\apj},
         year = 2006,
        month = dec,
       volume = {653},
       number = {2},
        pages = {1510-1516},
          doi = {10.1086/508565},
       adsurl = {https://ui.adsabs.harvard.edu/abs/2006ApJ...653.1510R}
}

@ARTICLE{Cranmer2023,
       author = {{Cranmer}, Steven R. and {Chhiber}, Rohit and {Gilly}, Chris R. and {Cairns}, Iver H. and {Colaninno}, Robin C. and {McComas}, David J. and {Raouafi}, Nour E. and {Usmanov}, Arcadi V. and {Gibson}, Sarah E. and {DeForest}, Craig E.},
        title = "{The Sun's Alfv{\'e}n Surface: Recent Insights and Prospects for the Polarimeter to Unify the Corona and Heliosphere (PUNCH)}",
      journal = {\solphys},
         year = 2023,
        month = nov,
       volume = {298},
       number = {11},
          eid = {126},
        pages = {126},
          doi = {10.1007/s11207-023-02218-2},
archivePrefix = {arXiv},
       eprint = {2310.05887},
 primaryClass = {astro-ph.SR},
       adsurl = {https://ui.adsabs.harvard.edu/abs/2023SoPh..298..126C}
}

@ARTICLE{ViallBorovsky2020,
       author = {{Viall}, Nicholeen M. and {Borovsky}, Joseph E.},
        title = "{Nine Outstanding Questions of Solar Wind Physics}",
      journal = {Journal of Geophysical Research (Space Physics)},
         year = 2020,
        month = jul,
       volume = {125},
       number = {7},
          eid = {e26005},
        pages = {e26005},
          doi = {10.1029/2018JA02600510.1002/essoar.10502606.1},
       adsurl = {https://ui.adsabs.harvard.edu/abs/2020JGRA..12526005V}
}

@ARTICLE{Jacques1977,
       author = {{Jacques}, S.~A.},
        title = "{Momentum and energy transport by waves in the solar atmosphere and solar wind.}",
      journal = {\apj},
         year = 1977,
        month = aug,
       volume = {215},
        pages = {942-951},
          doi = {10.1086/155430},
       adsurl = {https://ui.adsabs.harvard.edu/abs/1977ApJ...215..942J}
}

@ARTICLE{Downs2025,
       author = {{Downs}, Cooper and {Linker}, Jon A. and {Caplan}, Ronald M. and {Mason}, Emily I. and {Riley}, Pete and {Davidson}, Ryder and {Reyes}, Andres and {Palmerio}, Erika and {Lionello}, Roberto and {Turtle}, James and {Ben-Nun}, Michal and {Stulajter}, Miko M. and {Titov}, Viacheslav S. and {T{\"o}r{\"o}k}, Tibor and {Upton}, Lisa A. and {Attie}, Raphael and {Jha}, Bibhuti K. and {Arge}, Charles N. and {Henney}, Carl J. and {Valori}, Gherardo and {Strecker}, Hanna and {Calchetti}, Daniele and {Germerott}, Dietmar and {Hirzberger}, Johann and {Su{\'a}rez}, David Orozco and {Rodr{\'\i}guez}, Julian Blanco and {Solanki}, Sami K. and {Cheng}, Xin and {Wu}, Sizhe},
        title = "{A near-real-time data-assimilative model of the solar corona}",
      journal = {Science},
         year = 2025,
        month = jun,
       volume = {388},
       number = {6753},
        pages = {1306-1310},
          doi = {10.1126/science.adq0872},
       adsurl = {https://ui.adsabs.harvard.edu/abs/2025Sci...388.1306D}
}

@ARTICLE{Mason2023,
       author = {{Mason}, Emily I. and {Lionello}, Roberto and {Downs}, Cooper and {Linker}, Jon A. and {Caplan}, Ronald M. and {DeRosa}, Marc L.},
        title = "{Time-dependent Dynamics of the Corona}",
      journal = {\apjl},
         year = 2023,
        month = dec,
       volume = {959},
       number = {1},
          eid = {L4},
        pages = {L4},
          doi = {10.3847/2041-8213/ad00bd},
archivePrefix = {arXiv},
       eprint = {2306.11956},
 primaryClass = {astro-ph.SR},
       adsurl = {https://ui.adsabs.harvard.edu/abs/2023ApJ...959L...4M}
}

@ARTICLE{Milic2024,
       author = {{Mili{\'c}}, I. and {Centeno}, R. and {Sun}, X. and {Rempel}, M. and {de la Cruz Rodr{\'\i}guez}, J.},
        title = "{Spatial resolution effects on the solar open flux estimates}",
      journal = {\aap},
         year = 2024,
        month = mar,
       volume = {683},
          eid = {A134},
        pages = {A134},
          doi = {10.1051/0004-6361/202347272},
archivePrefix = {arXiv},
       eprint = {2402.02486},
 primaryClass = {astro-ph.SR},
       adsurl = {https://ui.adsabs.harvard.edu/abs/2024A&A...683A.134M}
}

@ARTICLE{delToroIniesta2016,
       author = {{del Toro Iniesta}, Jose Carlos and {Ruiz Cobo}, Basilio},
        title = "{Inversion of the radiative transfer equation for polarized light}",
      journal = {Living Reviews in Solar Physics},
         year = 2016,
        month = dec,
       volume = {13},
       number = {1},
          eid = {4},
        pages = {4},
          doi = {10.1007/s41116-016-0005-2},
archivePrefix = {arXiv},
       eprint = {1610.10039},
 primaryClass = {astro-ph.SR},
       adsurl = {https://ui.adsabs.harvard.edu/abs/2016LRSP...13....4D}
}

@ARTICLE{Klimchuk2006,
       author = {{Klimchuk}, James A.},
        title = "{On Solving the Coronal Heating Problem}",
      journal = {\solphys},
         year = 2006,
        month = mar,
       volume = {234},
       number = {1},
        pages = {41-77},
          doi = {10.1007/s11207-006-0055-z},
archivePrefix = {arXiv},
       eprint = {astro-ph/0511841},
 primaryClass = {astro-ph},
       adsurl = {https://ui.adsabs.harvard.edu/abs/2006SoPh..234...41K}
}

@ARTICLE{Aly1991,
       author = {{Aly}, J.~J.},
        title = "{How Much Energy Can Be Stored in a Three-dimensional Force-free Magnetic Field?}",
      journal = {\apjl},
         year = 1991,
        month = jul,
       volume = {375},
        pages = {L61},
          doi = {10.1086/186088},
       adsurl = {https://ui.adsabs.harvard.edu/abs/1991ApJ...375L..61A}
}

@ARTICLE{Sturrock1991,
       author = {{Sturrock}, P.~A.},
        title = "{Maximum Energy of Semi-infinite Magnetic Field Configurations}",
      journal = {\apj},
         year = 1991,
        month = oct,
       volume = {380},
        pages = {655},
          doi = {10.1086/170620},
       adsurl = {https://ui.adsabs.harvard.edu/abs/1991ApJ...380..655S}
}

@ARTICLE{Centeno2023,
       author = {{Centeno}, Rebecca and {Mili{\'c}}, Ivan and {Rempel}, Matthias and {Nitta}, Nariaki V. and {Sun}, Xudong},
        title = "{Limitations and Biases in the Retrieval of the Polar Magnetic Field. I. The Role of the Magnetic Filling Factor in Milne-Eddington Inversions of Simulated Hinode/SP Data}",
      journal = {\apj},
         year = 2023,
        month = jul,
       volume = {951},
       number = {1},
          eid = {23},
        pages = {23},
          doi = {10.3847/1538-4357/acd178},
archivePrefix = {arXiv},
       eprint = {2305.00924},
 primaryClass = {astro-ph.SR},
       adsurl = {https://ui.adsabs.harvard.edu/abs/2023ApJ...951...23C}
}

@ARTICLE{Titov2025,
       author = {{Titov}, Viacheslav S. and {Downs}, Cooper and {T{\"o}r{\"o}k}, Tibor and {Linker}, Jon A. and {Prazak}, Michael and {Qiu}, Jiong A.},
        title = "{Magnetogram-matching Biot{\textendash}Savart Law and Decomposition of Vector Magnetograms}",
      journal = {\apj},
         year = 2025,
        month = jul,
       volume = {988},
       number = {1},
          eid = {98},
        pages = {98},
          doi = {10.3847/1538-4357/add895},
archivePrefix = {arXiv},
       eprint = {2410.02134},
 primaryClass = {astro-ph.SR},
       adsurl = {https://ui.adsabs.harvard.edu/abs/2025ApJ...988...98T}
}

@ARTICLE{Schuck2022,
       author = {{Schuck}, Peter W. and {Linton}, Mark G. and {Knizhnik}, Kalman J. and {Leake}, James E.},
        title = "{On the Origin of the Photospheric Magnetic Field}",
      journal = {\apj},
         year = 2022,
        month = sep,
       volume = {936},
       number = {1},
          eid = {94},
        pages = {94},
          doi = {10.3847/1538-4357/ac739a},
       adsurl = {https://ui.adsabs.harvard.edu/abs/2022ApJ...936...94S}
}

@ARTICLE{Evans2026,
       author = {{Evans}, Caroline L. and {Downs}, Cooper and {Schmit}, Donald},
        title = "{Quantifying How Surface Complexity Influences Properties of the Solar Corona and Solar Wind}",
      journal = {\apj},
         year = 2026,
        month = jan,
       volume = {997},
       number = {1},
          eid = {80},
        pages = {80},
          doi = {10.3847/1538-4357/ae197a},
       adsurl = {https://ui.adsabs.harvard.edu/abs/2026ApJ...997...80E}
}

@ARTICLE{Mok2016,
       author = {{Mok}, Yung and {Miki{\'c}}, Zoran and {Lionello}, Roberto and {Downs}, Cooper and {Linker}, Jon A.},
        title = "{A Three-dimensional Model of Active Region 7986: Comparison of Simulations with Observations}",
      journal = {\apj},
         year = 2016,
        month = jan,
       volume = {817},
       number = {1},
          eid = {15},
        pages = {15},
          doi = {10.3847/0004-637X/817/1/15},
       adsurl = {https://ui.adsabs.harvard.edu/abs/2016ApJ...817...15M}
}

@ARTICLE{AnzerHeinzel2005,
       author = {{Anzer}, U. and {Heinzel}, P.},
        title = "{On the Nature of Dark Extreme Ultraviolet Structures Seen by SOHO/EIT and TRACE}",
      journal = {\apj},
         year = 2005,
        month = mar,
       volume = {622},
       number = {1},
        pages = {714-721},
          doi = {10.1086/427817},
       adsurl = {https://ui.adsabs.harvard.edu/abs/2005ApJ...622..714A}
}

@ARTICLE{Wang2009,
       author = {{Wang}, Zhou and {Bovik}, Alan C.},
        title = "{Mean Squared Error: Love it or Leave it?: A New Look at Signal Fidelity Measures}",
      journal = {IEEE Signal Processing Magazine},
         year = 2009,
        month = jan,
       volume = {26},
       number = {1},
        pages = {98-117},
          doi = {10.1109/MSP.2008.930649},
       adsurl = {https://ui.adsabs.harvard.edu/abs/2009ISPM...26...98W}
}

@ARTICLE{Wang2004,
       author = {{Wang}, Zhou and {Bovik}, Alan Conrad and {Sheikh}, Hamid Rahim and {Simoncelli}, Eero P.},
        title = "{Image Quality Assessment: From Error Visibility to Structural Similarity}",
      journal = {IEEE Transactions on Image Processing},
         year = 2004,
        month = jan,
       volume = {13},
       number = {4},
        pages = {600-612},
          doi = {10.1109/TIP.2003.819861},
       adsurl = {https://ui.adsabs.harvard.edu/abs/2004ITIP...13..600W}
}

@ARTICLE{Jarolim2025,
       author = {{Jarolim}, R. and {Veronig}, A.~M. and {P{\"o}tzi}, W. and {Podladchikova}, T.},
        title = "{A deep learning framework for instrument-to-instrument translation of solar observation data}",
      journal = {Nature Communications},
         year = 2025,
        month = apr,
       volume = {16},
       number = {1},
          eid = {3157},
        pages = {3157},
          doi = {10.1038/s41467-025-58391-4},
archivePrefix = {arXiv},
       eprint = {2401.08057},
 primaryClass = {astro-ph.SR},
       adsurl = {https://ui.adsabs.harvard.edu/abs/2025NatCo..16.3157J}
}

@ARTICLE{Schirninger2025,
       author = {{Schirninger}, C. and {Jarolim}, R. and {Veronig}, A.~M. and {Kuckein}, C.},
        title = "{Deep learning image burst stacking to reconstruct high-resolution ground-based solar observations}",
      journal = {\aap},
         year = 2025,
        month = jan,
       volume = {693},
          eid = {A6},
        pages = {A6},
          doi = {10.1051/0004-6361/202451850},
archivePrefix = {arXiv},
       eprint = {2506.04781},
 primaryClass = {astro-ph.SR},
       adsurl = {https://ui.adsabs.harvard.edu/abs/2025A&A...693A...6S}
}

@ARTICLE{Jarolim2024,
       author = {{Jarolim}, Robert and {Tremblay}, Benoit and {Mu{\~n}oz-Jaramillo}, Andr{\'e}s and {Bintsi}, Kyriaki-Margarita and {Jungbluth}, Anna and {Santos}, Miraflor and {Vourlidas}, Angelos and {Mason}, James P. and {Sundaresan}, Sairam and {Downs}, Cooper and {Caplan}, Ronald M.},
        title = "{SuNeRF: 3D Reconstruction of the Solar EUV Corona Using Neural Radiance Fields}",
      journal = {\apjl},
         year = 2024,
        month = feb,
       volume = {961},
       number = {2},
          eid = {L31},
        pages = {L31},
          doi = {10.3847/2041-8213/ad12d2},
archivePrefix = {arXiv},
       eprint = {2401.16388},
 primaryClass = {astro-ph.SR},
       adsurl = {https://ui.adsabs.harvard.edu/abs/2024ApJ...961L..31J}
}

@ARTICLE{Grajeda2025,
       author = {{Grajeda}, Jeremy A. and {Boucheron}, Laura E. and {Kirk}, Michael S. and {Leisner}, Andrew and {Landeros}, Jaime A. and {Arge}, C. Nick},
        title = "{Incorporating Magnetic-Field Characteristics into EUV-Based Automated Segmentation of Coronal Holes}",
      journal = {\solphys},
         year = 2025,
        month = sep,
       volume = {300},
       number = {9},
          eid = {126},
        pages = {126},
          doi = {10.1007/s11207-025-02536-7},
archivePrefix = {arXiv},
       eprint = {2501.13211},
 primaryClass = {astro-ph.SR},
       adsurl = {https://ui.adsabs.harvard.edu/abs/2025SoPh..300..126G}
}

@ARTICLE{Lemen2012,
       author = {{Lemen}, James R. and {Title}, Alan M. and {Akin}, David J. and {Boerner}, Paul F. and {Chou}, Catherine and {Drake}, Jerry F. and {Duncan}, Dexter W. and {Edwards}, Christopher G. and {Friedlaender}, Frank M. and {Heyman}, Gary F. and {Hurlburt}, Neal E. and {Katz}, Noah L. and {Kushner}, Gary D. and {Levay}, Michael and {Lindgren}, Russell W. and {Mathur}, Dnyanesh P. and {McFeaters}, Edward L. and {Mitchell}, Sarah and {Rehse}, Roger A. and {Schrijver}, Carolus J. and {Springer}, Larry A. and {Stern}, Robert A. and {Tarbell}, Theodore D. and {Wuelser}, Jean-Pierre and {Wolfson}, C. Jacob and {Yanari}, Carl and {Bookbinder}, Jay A. and {Cheimets}, Peter N. and {Caldwell}, David and {Deluca}, Edward E. and {Gates}, Richard and {Golub}, Leon and {Park}, Sang and {Podgorski}, William A. and {Bush}, Rock I. and {Scherrer}, Philip H. and {Gummin}, Mark A. and {Smith}, Peter and {Auker}, Gary and {Jerram}, Paul and {Pool}, Peter and {Soufli}, Regina and {Windt}, David L. and {Beardsley}, Sarah and {Clapp}, Matthew and {Lang}, James and {Waltham}, Nicholas},
        title = "{The Atmospheric Imaging Assembly (AIA) on the Solar Dynamics Observatory (SDO)}",
      journal = {\solphys},
         year = 2012,
        month = jan,
       volume = {275},
       number = {1-2},
        pages = {17-40},
          doi = {10.1007/s11207-011-9776-8},
       adsurl = {https://ui.adsabs.harvard.edu/abs/2012SoPh..275...17L}
}

@ARTICLE{ODwyer2010,
       author = {{O'Dwyer}, B. and {Del Zanna}, G. and {Mason}, H.~E. and {Weber}, M.~A. and {Tripathi}, D.},
        title = "{SDO/AIA response to coronal hole, quiet Sun, active region, and flare plasma}",
      journal = {\aap},
         year = 2010,
        month = oct,
       volume = {521},
          eid = {A21},
        pages = {A21},
          doi = {10.1051/0004-6361/201014872},
       adsurl = {https://ui.adsabs.harvard.edu/abs/2010A&A...521A..21O}
}

@ARTICLE{Morton2025,
       author = {{Morton}, R.~J. and {Weberg}, M.~J. and {Balodhi}, N. and {McLaughlin}, J.~A.},
        title = "{Estimating the Poynting Flux of Alfv{\'e}nic Waves in Polar Coronal Holes across Solar Cycle 24}",
      journal = {\apj},
         year = 2025,
        month = may,
       volume = {985},
       number = {1},
          eid = {13},
        pages = {13},
          doi = {10.3847/1538-4357/adc568},
archivePrefix = {arXiv},
       eprint = {2501.13673},
 primaryClass = {astro-ph.SR},
       adsurl = {https://ui.adsabs.harvard.edu/abs/2025ApJ...985...13M}
}

@ARTICLE{Wang2017,
       author = {{Wang}, Y.-M.},
        title = "{Small Coronal Holes Near Active Regions as Sources of Slow Solar Wind}",
      journal = {\apj},
         year = 2017,
        month = jun,
       volume = {841},
       number = {2},
          eid = {94},
        pages = {94},
          doi = {10.3847/1538-4357/aa706e},
       adsurl = {https://ui.adsabs.harvard.edu/abs/2017ApJ...841...94W}
}

@ARTICLE{Cranmer2009,
       author = {{Cranmer}, Steven R.},
        title = "{Coronal Holes}",
      journal = {Living Reviews in Solar Physics},
         year = 2009,
        month = dec,
       volume = {6},
       number = {1},
          eid = {3},
        pages = {3},
          doi = {10.12942/lrsp-2009-3},
archivePrefix = {arXiv},
       eprint = {0909.2847},
 primaryClass = {astro-ph.SR},
       adsurl = {https://ui.adsabs.harvard.edu/abs/2009LRSP....6....3C}
}

@ARTICLE{Caplan2023,
       author = {{Caplan}, Ronald M. and {Mason}, Emily I. and {Downs}, Cooper and {Linker}, Jon A.},
        title = "{Improving Coronal Hole Detections and Open Flux Estimates}",
      journal = {\apj},
         year = 2023,
        month = nov,
       volume = {958},
       number = {1},
          eid = {43},
        pages = {43},
          doi = {10.3847/1538-4357/ad01b6},
       adsurl = {https://ui.adsabs.harvard.edu/abs/2023ApJ...958...43C}
}

@ARTICLE{HowardTappin2009,
       author = {{Howard}, Timothy A. and {Tappin}, S. James},
        title = "{Interplanetary Coronal Mass Ejections Observed in the Heliosphere: 1. Review of Theory}",
      journal = {\ssr},
         year = 2009,
        month = oct,
       volume = {147},
       number = {1-2},
        pages = {31-54},
          doi = {10.1007/s11214-009-9542-5},
       adsurl = {https://ui.adsabs.harvard.edu/abs/2009SSRv..147...31H}
}

@ARTICLE{Downs2010,
       author = {{Downs}, Cooper and {Roussev}, Ilia I. and {van der Holst}, Bart and {Lugaz}, No{\'e} and {Sokolov}, Igor V. and {Gombosi}, Tamas I.},
        title = "{Toward a Realistic Thermodynamic Magnetohydrodynamic Model of the Global Solar Corona}",
      journal = {\apj},
         year = 2010,
        month = apr,
       volume = {712},
       number = {2},
        pages = {1219-1231},
          doi = {10.1088/0004-637X/712/2/1219},
archivePrefix = {arXiv},
       eprint = {0912.2647},
 primaryClass = {astro-ph.SR},
       adsurl = {https://ui.adsabs.harvard.edu/abs/2010ApJ...712.1219D}
}

@ARTICLE{SchrijverMcMullen2000,
       author = {{Schrijver}, Carolus J. and {McMullen}, Rebecca A.},
        title = "{A Case for Resonant Scattering in the Quiet Solar Corona in Extreme-Ultraviolet Lines with High Oscillator Strengths}",
      journal = {\apj},
         year = 2000,
        month = mar,
       volume = {531},
       number = {2},
        pages = {1121-1128},
          doi = {10.1086/308497},
       adsurl = {https://ui.adsabs.harvard.edu/abs/2000ApJ...531.1121S}
}

@ARTICLE{WoodRaymond2000,
       author = {{Wood}, Kenneth and {Raymond}, John},
        title = "{Resonant Scattering of Emission Lines in Coronal Loops: Effects on Image Morphology and Line Ratios}",
      journal = {\apj},
         year = 2000,
        month = sep,
       volume = {540},
       number = {1},
        pages = {563-571},
          doi = {10.1086/309336},
archivePrefix = {arXiv},
       eprint = {astro-ph/0004131},
 primaryClass = {astro-ph},
       adsurl = {https://ui.adsabs.harvard.edu/abs/2000ApJ...540..563W}
}

@ARTICLE{Leisner2026,
       author = {{Leisner}, Andrew and {Arge}, C. Nick and {Liu}, Yang and {Antiochos}, Spiro K. and {Wallace}, Samantha and {Zhang}, Jie},
        title = "{The Effect of Spatial Resolution on the Wang─Sheeley─Arge Model Unsigned Open Flux Prediction}",
      journal = {\apj},
         year = 2026,
        month = mar,
       volume = {1000},
       number = {1},
          eid = {113},
        pages = {113},
          doi = {10.3847/1538-4357/ae45a0},
       adsurl = {https://ui.adsabs.harvard.edu/abs/2026ApJ..1000..113L}
}

@ARTICLE{Reiss2021,
       author = {{Reiss}, Martin A. and {Muglach}, Karin and {M{\"o}stl}, Christian and {Arge}, Charles N. and {Bailey}, Rachel and {Delouille}, V{\'e}ronique and {Garton}, Tadhg M. and {Hamada}, Amr and {Hofmeister}, Stefan and {Illarionov}, Egor and {Jarolim}, Robert and {Kirk}, Michael S.~F. and {Kosovichev}, Alexander and {Krista}, Larisza and {Lee}, Sangwoo and {Lowder}, Chris and {MacNeice}, Peter J. and {Veronig}, Astrid and {Cospar Iswat Coronal Hole Boundary Working Team}},
        title = "{The Observational Uncertainty of Coronal Hole Boundaries in Automated Detection Schemes}",
      journal = {\apj},
         year = 2021,
        month = may,
       volume = {913},
       number = {1},
          eid = {28},
        pages = {28},
          doi = {10.3847/1538-4357/abf2c8},
archivePrefix = {arXiv},
       eprint = {2103.14403},
 primaryClass = {astro-ph.SR},
       adsurl = {https://ui.adsabs.harvard.edu/abs/2021ApJ...913...28R}
}

@ARTICLE{SunPy2020,
       author = {{SunPy Community} and {Barnes}, Will T. and {Bobra}, Monica G. and {Christe}, Steven D. and {Freij}, Nabil and {Hayes}, Laura A. and {Ireland}, Jack and {Mumford}, Stuart and {Perez-Suarez}, David and {Ryan}, Daniel F. and {Shih}, Albert Y. and {Chanda}, Prateek and {Glogowski}, Kolja and {Hewett}, Russell and {Hughitt}, V. Keith and {Hill}, Andrew and {Hiware}, Kaustubh and {Inglis}, Andrew and {Kirk}, Michael S.~F. and {Konge}, Sudarshan and {Mason}, James Paul and {Maloney}, Shane Anthony and {Murray}, Sophie A. and {Panda}, Asish and {Park}, Jongyeob and {Pereira}, Tiago M.~D. and {Reardon}, Kevin and {Savage}, Sabrina and {Sip{\H{o}}cz}, Brigitta M. and {Stansby}, David and {Jain}, Yash and {Taylor}, Garrison and {Yadav}, Tannmay and {Rajul} and {Dang}, Trung Kien},
        title = "{The SunPy Project: Open Source Development and Status of the Version 1.0 Core Package}",
      journal = {\apj},
         year = 2020,
        month = feb,
       volume = {890},
       number = {1},
          eid = {68},
        pages = {68},
          doi = {10.3847/1538-4357/ab4f7a},
       adsurl = {https://ui.adsabs.harvard.edu/abs/2020ApJ...890...68S}
}

@ARTICLE{Petrie2025,
       author = {{Petrie}, Gordon J.~D. and {Blanco Rodr{\'\i}guez}, Juli{\'a}n and {Mart{\'\i}nez Pillet}, Valent{\'\i}n and {Uitenbroek}, Han and {Scherrer}, Philip H.},
        title = "{Calibrating HMI Magnetograms Using an End-to-end Magnetograph Model}",
      journal = {\apjs},
         year = 2025,
        month = jun,
       volume = {278},
       number = {2},
          eid = {55},
        pages = {55},
          doi = {10.3847/1538-4365/adcf9f},
       adsurl = {https://ui.adsabs.harvard.edu/abs/2025ApJS..278...55P}
}

@ARTICLE{Liu2026,
       author = {{Liu}, Xianyu and {Liu}, Weihao and {Manchester}, IV, Ward B. and {Welling}, Daniel T. and {T{\'o}th}, G{\'a}bor and {Gombosi}, Tamas I. and {DeRosa}, Marc L. and {Bertello}, Luca and {Pevtsov}, Alexei A. and {Pevtsov}, Alexander A. and {Reardon}, Kevin and {Wilbanks}, Kathryn and {Rewoldt}, Amy and {Zhao}, Lulu},
        title = "{Simulating the Solar Corona with Multiple Solar Photospheric Magnetic Maps during the 2024 April 8 Total Solar Eclipse}",
      journal = {\apj},
         year = 2026,
        month = feb,
       volume = {997},
       number = {2},
          eid = {243},
        pages = {243},
          doi = {10.3847/1538-4357/ae290f},
       adsurl = {https://ui.adsabs.harvard.edu/abs/2026ApJ...997..243L}
}

@ARTICLE{Baratashvili2026,
       author = {{Baratashvili}, T. and {Wang}, H.~P. and {Sorokina}, D. and {Lani}, A. and {Poedts}, S.},
        title = "{Modelling the total solar eclipse in 2024 with COCONUT}",
      journal = {\aap},
         year = 2026,
        month = jan,
       volume = {705},
          eid = {A145},
        pages = {A145},
          doi = {10.1051/0004-6361/202556300},
archivePrefix = {arXiv},
       eprint = {2511.06535},
 primaryClass = {astro-ph.SR},
       adsurl = {https://ui.adsabs.harvard.edu/abs/2026A&A...705A.145B}
}

@ARTICLE{Sachdeva2026,
       author = {{Sachdeva}, Nishtha and {Huang}, Zhenguang and {Toth}, Gabor and {Chen}, Hongfan and {Manchester}, Ward B. and {van der Holst}, Bart},
        title = "{Evolution of Coronal Mass Ejections in Different Data-driven Solar Wind Conditions}",
      journal = {\apj},
         year = 2026,
        month = apr,
       volume = {1001},
       number = {2},
          eid = {144},
        pages = {144},
          doi = {10.3847/1538-4357/ae4348},
archivePrefix = {arXiv},
       eprint = {2602.02300},
 primaryClass = {astro-ph.SR},
       adsurl = {https://ui.adsabs.harvard.edu/abs/2026ApJ..1001..144S}
}

@ARTICLE{Huang2023,
       author = {{Huang}, Zhenguang and {T{\'o}th}, G{\'a}bor and {Sachdeva}, Nishtha and {Zhao}, Lulu and {van der Holst}, Bart and {Sokolov}, Igor and {Manchester}, Ward B. and {Gombosi}, Tamas I.},
        title = "{Modeling the Solar Wind during Different Phases of the Last Solar Cycle}",
      journal = {\apjl},
         year = 2023,
        month = apr,
       volume = {946},
       number = {2},
          eid = {L47},
        pages = {L47},
          doi = {10.3847/2041-8213/acc5ef10.1002/essoar.10512539.1},
       adsurl = {https://ui.adsabs.harvard.edu/abs/2023ApJ...946L..47H}
}

@ARTICLE{Huang2024,
       author = {{Huang}, Zesen and {Velli}, Marco and {Shi}, Chen and {Zhu}, Yingjie and {Chandran}, B.~D.~G. and {Bowen}, Trevor and {R{\'e}ville}, Victor and {Huang}, Jia and {Hou}, Chuanpeng and {Sioulas}, Nikos and {Liu}, Mingzhe and {Pulupa}, Marc and {Huang}, Sheng and {Bale}, Stuart D.},
        title = "{Dominance of 2 Minute Oscillations near the Alfv{\'e}n Surface}",
      journal = {\apjl},
         year = 2024,
        month = dec,
       volume = {977},
       number = {1},
          eid = {L12},
        pages = {L12},
          doi = {10.3847/2041-8213/ad9271},
archivePrefix = {arXiv},
       eprint = {2405.15967},
 primaryClass = {astro-ph.SR},
       adsurl = {https://ui.adsabs.harvard.edu/abs/2024ApJ...977L..12H}
}

@ARTICLE{Riley2019,
       author = {{Riley}, Pete and {Linker}, Jon A. and {Mikic}, Zoran and {Caplan}, Ronald M. and {Downs}, Cooper and {Thumm}, Jean-Luc},
        title = "{Can an Unobserved Concentration of Magnetic Flux Above the Poles of the Sun Resolve the Open Flux Problem?}",
      journal = {\apj},
         year = 2019,
        month = oct,
       volume = {884},
       number = {1},
          eid = {18},
        pages = {18},
          doi = {10.3847/1538-4357/ab3a98},
       adsurl = {https://ui.adsabs.harvard.edu/abs/2019ApJ...884...18R}
}
\bibliographystyle{aasjournalv7}

\end{document}